\documentclass[aps,prl,onecolumn,10pt,superscriptaddress,amsmath,amssymb,floatfix,longbibliography]{revtex4-2}
\usepackage[T1]{fontenc}
\usepackage{array}[=2016-10-06]
\usepackage{booktabs,bm,microtype,mathrsfs,amsthm}
\usepackage{xcolor}
\usepackage{tikz}
\usetikzlibrary{arrows.meta}
\usetikzlibrary{arrows.meta}
\definecolor{cfOutline}{HTML}{3C5364}
\definecolor{cfReference}{HTML}{8B9BA8}
\definecolor{cfGrid}{HTML}{8899A6}
\definecolor{cfRed}{HTML}{9B4E51}
\definecolor{cfBlue}{HTML}{315F7D}
\definecolor{cfRedFill}{HTML}{F6E8E7}
\definecolor{cfBlueFill}{HTML}{E8F1F6}

\tikzset{
  cf panel/.style={x=12.7mm,y=12.7mm,line cap=round,line join=round},
  cf outline/.style={draw=cfOutline,line width=.90pt},
  cf reference/.style={draw=cfReference,line width=.76pt,dash pattern=on 2.5pt off 2.0pt},
  cf mesh/.style={draw=cfGrid,line width=.47pt,dash pattern=on 1.0pt off 1.3pt},
  cf motion/.style={draw=cfOutline,line width=1.07pt,-{Stealth[length=4.4pt,width=3.7pt]}},
  cf title/.style={font=\small,text=black,anchor=center,inner sep=0pt}
}

\newcommand{\cfPanelBounds}{%
  \path[use as bounding box] (-2.83,-1.68) rectangle (2.83,1.11);%
}
\newcommand{\cfTitle}[2]{%
  \node[cf title] at (0,-1.34) {\(\mathbf{#1}\)\enspace #2};%
}

\pgfmathdeclarefunction{cfradial}{1}{\pgfmathparse{sqrt(max(0,1-(#1)^2))}}
\pgfmathdeclarefunction{cfx}{2}{\pgfmathparse{#1*cos(15)-cfradial(#1)*sin(15)*cos(#2)}}
\pgfmathdeclarefunction{cfy}{2}{\pgfmathparse{cfradial(#1)*sin(#2)}}

\newcommand{\cfShell}[3]{%
  \begin{scope}
    \clip #1;
    \fill[cfRedFill] (-1.25,-1.2) rectangle (#2,1.2);
    \fill[cfBlueFill] (#2,-1.2) rectangle (1.25,1.2);
    #3%
  \end{scope}
  \draw[cf outline] #1;%
}
\newcommand{\cfReferenceSphere}{\draw[cf reference] (0,0) circle[radius=1];}

\newcommand{\cfCrossSections}{%
  \foreach \cfU in {-.79,-.54,-.28,0,.28,.54,.79} {
    \pgfmathsetmacro{\cfAngle}{acos(max(-1,min(1,-\cfU*tan(15)/cfradial(\cfU))))}
    \draw[cf mesh] plot[domain=-\cfAngle:\cfAngle,samples=25,variable=\cfT]
      ({cfx(\cfU,\cfT)},{cfy(\cfU,\cfT)});
  }%
}
\newcommand{\cfSphereMesh}{%
  \foreach \cfAngle in {-143,-109,-75,-42,-13,18,50,82,113,147} {
    \pgfmathsetmacro{\cfStart}{max(-.989,-cos(\cfAngle)*cos(15)/
      sqrt(sin(15)^2+(cos(\cfAngle)*cos(15))^2))}
    \draw[cf mesh] plot[domain=\cfStart:.992,samples=25,variable=\cfT]
      ({cfx(\cfT,\cfAngle)},{cfy(\cfT,\cfAngle)});
  }
  \cfCrossSections
}
\newcommand{\cfVerticalMesh}{%
  \begin{scope}[cm={0,1,1,0,(0,0)}]
    \cfSphereMesh
  \end{scope}%
}

\newcommand{\cfDephasing}[2]{%
  \pgfmathsetmacro{\cfXX}{cos(#1)^2+#2*sin(#1)^2}
  \pgfmathsetmacro{\cfXY}{(1-#2)*sin(#1)*cos(#1)}
  \pgfmathsetmacro{\cfYY}{sin(#1)^2+#2*cos(#1)^2}
  \def\cfEllipse{plot[domain=0:360,samples=65,variable=\cfT]
    ({\cfXX*cos(\cfT)+\cfXY*sin(\cfT)},{\cfXY*cos(\cfT)+\cfYY*sin(\cfT)}) -- cycle}
  \cfReferenceSphere
  \cfShell{\cfEllipse}{0}{%
    \begin{scope}[cm={\cfXX,\cfXY,\cfXY,\cfYY,(0,0)}]
      \cfVerticalMesh
    \end{scope}
  }
  \draw[cf motion,cfRed] ({-sin(#1)},{cos(#1)}) -- ({-#2*sin(#1)},{#2*cos(#1)});
  \draw[cf motion,cfBlue] ({sin(#1)},{-cos(#1)}) -- ({#2*sin(#1)},{-#2*cos(#1)});
}
\newcolumntype{L}[1]{>{\raggedright\arraybackslash}p{#1}}
\usepackage[hidelinks]{hyperref}
\hypersetup{pdftitle={Modeling Noise Beyond Pauli Channels in Quantum Error Correction},pdflang={en}}
\newcommand{\id}{\mathcal I}
\newcommand{\C}{\mathcal C}
\newcommand{\N}{\mathcal N}
\newcommand{\tr}{\operatorname{tr}}
\newcommand{\proj}[1]{|#1\rangle\langle #1|}
\newtheorem{theorem}{Theorem}
\newtheorem{proposition}[theorem]{Proposition}
\theoremstyle{definition}
\newtheorem{definition}[theorem]{Definition}

\begin{document}
\title{Modeling Noise Beyond Pauli Channels\texorpdfstring{\\}{ }in Quantum Error Correction}

\newcommand{\unitaryForgeAddress}{UnitaryForge, Hangzhou 311200, China}

\author{Ximing Wang}
\affiliation{\unitaryForgeAddress}
\affiliation{Nanyang Technological University, Singapore 637371, Singapore}

\author{Chufan Lyu}
\affiliation{\unitaryForgeAddress}

\author{Chengran Yang}
\affiliation{\unitaryForgeAddress}
\affiliation{Nanyang Technological University, Singapore 637371, Singapore}

\begin{abstract}
Predicting logical errors requires a noise model that captures physical faults and their correlations across circuit operations.
We systematically construct such models by composing a finite number of elementary physical channels. The channels form six families, each with at most three continuous parameters for a fixed choice of Pauli axes. Their compositions can describe both non-Pauli noise and non-Markovian effects.
For example, the population-transfer family reproduces single-qubit amplitude damping. The same mechanism can bias the parity of a maximally mixed two-qubit state while leaving both single-qubit states maximally mixed.
To model non-Markovian noise on the data qubits, we retain the leakage occupation across circuit operations. Leakage at one location can then determine whether a neighboring qubit dephases at a later one.
This construction provides a new way to incorporate these physical mechanisms into circuit models and study their effects on logical errors in future simulations of quantum error correction.
\end{abstract}

\maketitle

\section{Introduction}
\label{sec:introduction}

Predicting how well a quantum error-correcting code protects information requires a model of the noise in the correction circuit. In simulations of a surface code with coherent physical errors, Pauli twirling underestimated the logical error rate below threshold~\cite{bravyi2018}. Noise models can also guide the choice of code and decoder for quantum error correction (QEC). A code can exploit a bias toward one type of Pauli error~\cite{bonilla2021}. A decoder selects a recovery operation from syndrome records~\cite{dennis2002} and can learn to interpret those records using examples generated by a noise model~\cite{bausch2024}.

Stochastic Pauli channels provide a simple starting point. Assigning probabilities to Pauli strings captures both biases in individual errors and correlations across qubits. These channels preserve the maximally mixed state. Relaxation can shift its populations, and coherent rotations can change relative phases. In a transmon controlled-$Z$ gate, leakage of one qubit can also change the phase rotation of its partner~\cite{varbanov2020}. These mechanisms motivate a common framework in which a few parameters have identifiable physical effects.

We construct six families of physical channels from elementary error generators in the Pauli basis~\cite{taxonomy2022,miller2025}. Two classes describe rotation and dephasing about a single axis. The four two-axis classes are distinguished by the commutation relation of the axes and the relative phase in the cross term. We complete each cross term with weighted dephasing along its two axes so that exponentiating the resulting generator gives a completely positive, trace-preserving channel. Once the axes are fixed, at most three continuous parameters specify a channel within each family. The channel formulas identify the population transferred between subspaces, the coherence that survives, and the operations that share a random event.

The choice of axes determines which degrees of freedom a mechanism affects. For one choice, the population-transfer family gives single-qubit amplitude damping. For another, it biases the parity of a maximally mixed two-qubit state while preserving the maximally mixed state of each qubit. We also combine the families in a four-level leakage model. A register that tracks the occupied subspace allows us to describe leakage and return, dephasing during leakage, and rotations of a neighboring qubit at transitions between the subspaces.

\section{Physical generators and channel dictionary}
\label{sec:foundations}

A Pauli axis divides the Hilbert space into two eigenspaces. The resulting block representation tracks the populations within the eigenspaces and the coherences between them. We use it to interpret the finite channel associated with each generator class.

\subsection{Pauli eigenspaces and the generator basis}
\label{sec:complete-jumps}

\begin{figure}[htbp]
\centering
\begingroup
\definecolor{psblue}{RGB}{38,79,110}
\begin{tikzpicture}[
  x=1.7cm,y=1.7cm,font=\small,
  psdirection/.style={draw=black!55,line width=.45pt,-{Stealth[length=1.4mm]}},
  pslabel/.style={inner sep=2pt},
  psspace/.style={font=\footnotesize,align=center,inner sep=3pt}
]
\draw[psdirection] (-1.26,0)--(1.40,0);
\draw[psdirection] (0,-1.22)--(0,1.34);

\draw[draw=psblue,line width=.95pt] (0,0)--(.6,0);
\draw[draw=psblue,line width=.95pt] (0,0)--(0,.8);
\draw[draw=black!50,densely dotted,line width=.5pt]
  (.6,0)--(.6,.8)--(0,.8);
\fill[psblue] (.6,0) circle (1.05pt);
\fill[psblue] (0,.8) circle (1.05pt);

\draw[draw=black,line width=.7pt,-{Stealth[length=1.7mm]}]
  (0,0)--(.6,.8);
\draw[draw=psblue,densely dashed,line width=.7pt,-{Stealth[length=1.7mm]}]
  (0,0)--(.6,-.8);
\fill[black] (.6,.8) circle (1.05pt);
\fill[psblue] (.6,-.8) circle (1.05pt);
\draw[draw=psblue,line width=.65pt,-{Stealth[length=1.6mm]}]
  (.68,.79) .. controls (1.33,.62) and (1.33,-.62) .. (.68,-.79);

\node[psspace,anchor=west] at (1.44,0)
  {$\mathcal V_+$\\[-1pt]$\dim=2^{n-1}$};
\node[psspace,anchor=south] at (0,1.36)
  {$\mathcal V_-$\\[-1pt]$\dim=2^{n-1}$};
\node[pslabel,anchor=south west] at (.62,.82) {$\psi$};
\node[pslabel,anchor=north west,text=psblue] at (.64,-.84) {$T\psi$};
\node[pslabel,anchor=north,text=psblue] at (.60,-.12) {$\Pi_+\psi$};
\node[pslabel,anchor=east,text=psblue] at (-.10,.60) {$\Pi_-\psi$};
\end{tikzpicture}
\endgroup
\caption{Reflection about a Pauli eigenspace. Dotted guides and highlighted segments identify the components $\Pi_\pm\psi$ of the solid vector $\psi$ in $\mathcal V_\pm$. The operator $T$ preserves the component in $\mathcal V_+$ and reverses the one in $\mathcal V_-$, yielding the dashed vector $T\psi=\Pi_+\psi-\Pi_-\psi$. The curved arrow indicates the reflection.}
\label{fig:pauli-sectors}
\end{figure}
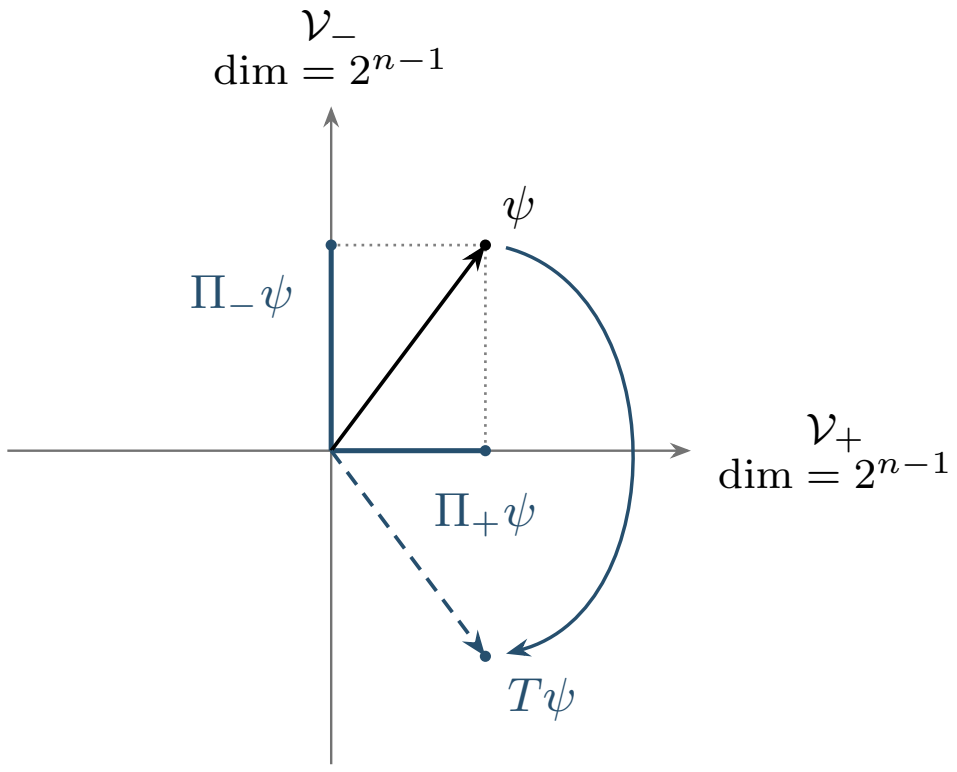

Let $\mathcal P_n=\{I,X,Y,Z\}^{\otimes n}$ be the Pauli strings on $n$ qubits.
The $4^n$ Hermitian strings, including the identity $I$, form an orthogonal basis for the operator space.
We define a nonidentity Pauli axis $T$ by choosing a nonidentity string and a sign.
Its eigenspaces $\mathcal V_+$ and $\mathcal V_-$ have eigenvalues $+1$ and $-1$, respectively, and partition $\mathcal H=(\mathbb C^2)^{\otimes n}$ into orthogonal subspaces of dimension $2^{n-1}$.
With the corresponding projectors $\Pi_\pm=(I\pm T)/2$, the density operator takes the block form
\begin{equation}
\rho=\begin{pmatrix}
\rho_{++}&\rho_{+-}\\
\rho_{-+}&\rho_{--}
\end{pmatrix},
\qquad \rho_{\sigma\tau}=\Pi_\sigma\rho\Pi_\tau,
\qquad \sigma,\tau\in\{+,-\}.
\label{eq:rho-sector-blocks}
\end{equation}
Each diagonal block describes the unnormalized state in one eigenspace, with $\operatorname{tr}\rho_{\tau\tau}$ giving the population of $T$'s eigenspace. The off-diagonal blocks describe coherence between the eigenspaces.
Figure~\ref{fig:pauli-sectors} illustrates the action of $T$ on the corresponding components of a state vector.

A superoperator is a complex-linear map on operators. A quantum channel is a completely positive, trace-preserving (CPTP) superoperator, and the identity channel acts as $\mathcal I(X)=X$.
We describe perturbations about $\mathcal I$ using trace-annihilating, Hermiticity-preserving (TAHP) maps. These maps send Hermitian inputs to Hermitian outputs and have zero output trace for every input.
For a TAHP map $\mathcal L$, the exponential $e^{\mathcal L}$ preserves trace and Hermiticity and approaches $\mathcal I$ as $\|\mathcal L\|$ tends to zero. It is a physical channel if it is completely positive.


We express the generator directions in Pauli coordinates by defining
\begin{equation}
\label{eq:pauli-axis-map}
\mathcal F_{PQ}(\alpha)(X)
=i^\alpha PXQ+i^{-\alpha}QXP
-\tfrac12\{i^\alpha QP+i^{-\alpha}PQ,X\} \ ,
\end{equation}
where $P,Q\in\mathcal P_n$ and $\alpha\in\{0,1\}$. The same formula applies to signed Pauli strings.
The resulting TAHP maps span the real vector space of TAHP superoperators.

Fix a total order $<$ on the nonidentity Pauli axes. A real basis consists of $\mathcal F_{IP}(1)$ and $\mathcal F_{PP}(0)$ for each $P\ne I$, and $\mathcal F_{PQ}(0)$ and $\mathcal F_{PQ}(1)$ for distinct nonidentity axes with $P<Q$~\cite{taxonomy2022}.
The identity pair gives $\mathcal F_{II}(\alpha)=0$.
For two distinct nonidentity strings, the commutation relation and the choice of $\alpha$ distinguish four cross-term classes. Including the single-axis rotation and dephasing classes gives the six classes in Table~\ref{tab:generator-classes}~\cite{taxonomy2022}.

\begin{table}[htbp]
\caption{From Pauli pairs to generator classes. For a cross term, $P$ and $Q$ are distinct nonidentity Pauli strings.
The physical effect in the last column refers to the full generator in Table~\ref{tab:physical-families}.}
\label{tab:generator-classes}
\centering\small
\setlength{\tabcolsep}{5pt}
\renewcommand{\arraystretch}{1.2}
\begin{tabular}{@{}L{0.34\linewidth}L{0.22\linewidth}L{0.38\linewidth}@{}}
\toprule
Class from $\mathcal F$ & Axes & Physical association \\
\midrule
$\mathbf{H}_P=\mathcal F_{IP}(1)$ & $P\ne I$ & Coherent Pauli rotation \\
$\mathbf{S}_P=\mathcal F_{PP}(0)$ & $P\ne I$ & Pauli dephasing \\
$\mathbf{C}_{P,Q}=\mathcal F_{PQ}(0)$ & $[P,Q]=0$ & Subspace-dependent dephasing \\
$\mathbf{A}_{P,Q}=\mathcal F_{PQ}(1)$ & $\{P,Q\}=0$ & Population transfer between subspaces \\
$\mathbf{E}_{P,Q}=\mathcal F_{PQ}(1)$ & $[P,Q]=0$ & Event-correlated phase kicks \\
$\mathbf{F}_{P,Q}=\mathcal F_{PQ}(0)$ & $\{P,Q\}=0$ & Rotated-axis dephasing \\
\bottomrule
\end{tabular}
\end{table}

\subsection{Six exponential families}
\label{sec:families}
\label{sec:conversions}
\label{sec:damping-threshold}

We define six families of generators, whose channel actions are illustrated in table~\ref{tab:physical-families}. Their exponentials are physical channels for $s,\lambda\geq 0$ and $\theta\in\mathbb R$. A fixed generator gives Markovian evolution on the full register. The data alone can exhibit memory if a leakage degree of freedom persists between circuit locations.

Rotation ($\mathbf{H}$) and dephasing ($\mathbf{S}$) each use a single Pauli axis $T$. For the other four families, we combine a cross term with dephasing along $P$ and $Q$, with relative weights set by $\theta$ according to
\begin{equation}
\label{eq:CrossTerm-generators}
G_\nu(\theta) = \tfrac12\bigl(\cos^2\theta\,\mathbf{S}_P + \sin^2\theta\,\mathbf{S}_Q + \sin2\theta\,\nu_{P,Q}\bigr) \ ,
\end{equation}
where $\nu\in\{\mathbf{C},\mathbf{A},\mathbf{E},\mathbf{F}\}$ denotes the cross-term class. We choose $T=PQ$ for commuting axes and $T=-iPQ$ for anticommuting axes.
The strength $s$ scales $G_\nu$. The parameter $\lambda$ sets dephasing along $T$ in the $\mathbf{S}$, $\mathbf{C}$ and $\mathbf{A}$ families, and $\theta$ sets the coherent rotation in the $\mathbf{H}$ family. We write $\mathcal C_U(X)=UXU^\dagger$ for conjugation by a unitary $U$.

\begin{table}[htbp]
\caption{Complete generators and illustrations for the six finite-strength channels.
Exponentiating each generator gives the named physical channel for $s,\lambda\geq 0$ and $\theta\in\mathbb R$. The four cross-term families use $G_\nu$ from Eq.~\eqref{eq:CrossTerm-generators}.
Dashed outlines show the undeformed sphere. The $\mathbf{A}$ sketch shows amplitude damping with the right pole fixed, $\mathbf{C}$ shows unequal dephasing within the two subspaces, and $\mathbf{E}$ shows their relative phase for a fixed event count.}
\label{tab:physical-families}
\centering\small
\setlength{\tabcolsep}{10pt}
\renewcommand{\arraystretch}{1.2}
\renewcommand{\cfPanelBounds}{%
  \path[use as bounding box] (-1.16,-1.16) rectangle (1.16,1.16);%
}
\renewcommand{\cfTitle}[2]{}
\tikzset{cf panel/.append style={baseline=-.12cm}}
\begin{tabular}{@{}llc@{}}
\toprule
Physical family & Complete generator & Illustration \\
\midrule
Coherent Pauli rotation ($\mathbf{H}$) & $\theta \mathbf{H}_T$ & \scalebox{0.58}{%
\begin{tikzpicture}[cf panel]
\cfPanelBounds
\cfShell{(0,0) circle[radius=1]}{0}{\cfVerticalMesh}
\fill[cfReference] ({cfy(.41,-68)},{cfx(.41,-68)}) circle[radius=1.9pt];
\draw[cf motion,line width=1.22pt] plot[domain=-68:61,samples=36,variable=\cfT]
  ({cfy(.41,\cfT)},{cfx(.41,\cfT)});
\cfTitle{H}{Coherent Pauli rotation}
\end{tikzpicture}%
} \\
Pauli dephasing ($\mathbf{S}$) & $\lambda \mathbf{S}_T$ & \scalebox{0.58}{%
\begin{tikzpicture}[cf panel]
\cfPanelBounds
\cfDephasing{90}{.45}
\cfTitle{S}{Pauli dephasing}
\end{tikzpicture}%
} \\
Subspace-dependent dephasing ($\mathbf{C}$) & $s G_{\mathbf{C}}(\theta)+\lambda \mathbf{S}_T$ & \scalebox{0.58}{%
\begin{tikzpicture}[cf panel]
\cfPanelBounds
\pgfmathdeclarefunction{cfattenuation}{1}{\pgfmathparse{.34+.50/(1+exp(-8*#1))}}
\pgfmathdeclarefunction{cfcompressedx}{2}{\pgfmathparse{#1-.163*(1-(#1)^2)*cos(#2)}}
\pgfmathdeclarefunction{cfcompressedy}{2}{\pgfmathparse{cfradial(#1)*cfattenuation(#1)*sin(#2)}}
\def\cfCompressedOutline{plot[domain=0:360,samples=97,variable=\cfT]
  ({cos(\cfT)},{sin(\cfT)*cfattenuation(cos(\cfT))}) -- cycle}
\cfReferenceSphere
\cfShell{\cfCompressedOutline}{0}{%
  \foreach \cfAngle in {-75,-44,-13,18,50,80} {
    \draw[cf mesh] plot[domain=-.985:.988,samples=31,variable=\cfT]
      ({cfcompressedx(\cfT,\cfAngle)},{cfcompressedy(\cfT,\cfAngle)});
  }
  \foreach \cfU in {-.79,-.54,-.28,0,.28,.54,.79} {
    \draw[cf mesh] plot[domain=-90:90,samples=23,variable=\cfT]
      ({cfcompressedx(\cfU,\cfT)},{cfcompressedy(\cfU,\cfT)});
  }
}
\foreach \cfU/\cfArrowColor in {-.51/cfRed,.49/cfBlue} {
  \foreach \cfSign in {-1,1} {
    \draw[cf motion,\cfArrowColor] (\cfU,{\cfSign*cfradial(\cfU)})
      -- (\cfU,{\cfSign*cfradial(\cfU)*cfattenuation(\cfU)});
  }
}
\cfTitle{C}{Subspace-dependent dephasing}
\end{tikzpicture}%
} \\
Population transfer between subspaces ($\mathbf{A}$) & $s G_{\mathbf{A}}(\theta)+\lambda \mathbf{S}_T$ & \scalebox{0.58}{%
\begin{tikzpicture}[cf panel]
\cfPanelBounds
\pgfmathsetmacro{\cfRetention}{.53}
\pgfmathsetmacro{\cfTransverse}{sqrt(\cfRetention)}
\pgfmathsetmacro{\cfShift}{1-\cfRetention}
\def\cfDampedOutline{(\cfShift,0) ellipse[x radius=\cfRetention,y radius=\cfTransverse]}
\cfReferenceSphere
\cfShell{\cfDampedOutline}{\cfShift}{%
  \begin{scope}[cm={\cfRetention,0,0,\cfTransverse,(\cfShift,0)}]
    \cfVerticalMesh
  \end{scope}
}
\foreach \cfAngle in {135,180,225} {
  \draw[cf motion,cfRed] (\cfAngle:1)
    -- ({\cfShift+\cfRetention*cos(\cfAngle)},{\cfTransverse*sin(\cfAngle)});
}
\fill[cfBlue] (1,0) circle[radius=2.05pt];
\cfTitle{A}{Population transfer between subspaces}
\end{tikzpicture}%
} \\
Event-correlated phase kicks ($\mathbf{E}$) & $s G_{\mathbf{E}}(\theta)$ & \scalebox{0.58}{%
\begin{tikzpicture}[cf panel]
\cfPanelBounds
\pgfmathdeclarefunction{cfwinding}{1}{\pgfmathparse{52.7-37.25*(2/(1+exp(-5*#1))-1)}}
\cfShell{(0,0) circle[radius=1]}{0}{%
  \foreach \cfAngle/\cfFirst/\cfLast in {
    -143/-.910/-.004,-143/.913/.988,-109/-.961/.988,-75/-.963/.988,
    -42/-.930/.988,-13/-.700/.988,18/-.188/.988,50/.120/.988,
    82/.588/.988,113/.921/.988,147/.963/.988} {
    \draw[cf mesh] plot[domain=\cfFirst:\cfLast,samples=31,variable=\cfT]
      ({cfx(\cfT,\cfAngle+cfwinding(\cfT))},{cfy(\cfT,\cfAngle+cfwinding(\cfT))});
  }
  \cfCrossSections
}
\foreach \cfU/\cfArrowColor in {-.49/cfRed,.56/cfBlue} {
  \pgfmathsetmacro{\cfEnd}{-37+cfwinding(\cfU)}
  \fill[cfReference] ({cfx(\cfU,-37)},{cfy(\cfU,-37)}) circle[radius=1.48pt];
  \draw[cf motion,\cfArrowColor] plot[domain=-37:\cfEnd,samples=27,variable=\cfT]
    ({cfx(\cfU,\cfT)},{cfy(\cfU,\cfT)});
}
\cfTitle{E}{Event-correlated phase kicks}
\end{tikzpicture}%
} \\
Rotated-axis dephasing ($\mathbf{F}$) & $s G_{\mathbf{F}}(\theta)$ & \scalebox{0.58}{%
\begin{tikzpicture}[cf panel]
\cfPanelBounds
\cfDephasing{48}{.45}
\cfTitle{F}{Rotated-axis dephasing}
\end{tikzpicture}%
} \\
\bottomrule
\end{tabular}
\end{table}

\emph{Coherent Pauli rotation}~($\mathbf{H}$).
A Pauli Hamiltonian imparts opposite phases to the two eigenspaces of $T$. The resulting rotation about $T$ is
\begin{equation}
\mathcal N_{\mathbf{H}}(\theta)=e^{\theta \mathbf{H}_T}=\mathcal C_{e^{-i\theta T}}.
\label{eq:coherentexp}
\end{equation}
In the block representation the relative phase appears only in the coherence between eigenspaces, giving
\begin{equation}
\mathcal N_{\mathbf{H}}(\theta)(\rho)=e^{-i\theta T}\rho e^{i\theta T}
=\begin{pmatrix}
\rho_{++}&e^{-2i\theta}\rho_{+-}\\[2pt]
e^{2i\theta}\rho_{-+}&\rho_{--}
\end{pmatrix}.
\label{eq:Hsectorblocks}
\end{equation}
Sequences of such Pauli rotations approximate small unitary errors by Trotterization~\cite{trotter1959}.

\emph{Pauli dephasing}~($\mathbf{S}$).
The $\mathbf{S}$ channel applies $T$ with probability $(1-\eta)/2$, reversing the relative sign of the two eigenspace components. It applies the identity otherwise. The resulting channel is
\begin{equation}
\mathcal N_{\mathbf{S}}(\lambda)=e^{\lambda \mathbf{S}_T}=\frac{1+\eta}2\,\mathcal I
          +\frac{1-\eta}2\,\mathcal C_T,
\qquad \eta=e^{-4\lambda}.
\label{eq:pauliexp}
\end{equation}
The random sign changes preserve the populations and leave a fraction $\eta$ of the coherence between eigenspaces. This gives the block action
\begin{equation}
\mathcal N_{\mathbf{S}}(\lambda)(\rho)=
\begin{pmatrix}
\rho_{++}&\eta\rho_{+-}\\
\eta\rho_{-+}&\rho_{--}
\end{pmatrix}.
\label{eq:Ssectorblocks}
\end{equation}

\emph{Subspace-dependent dephasing}~($\mathbf{C}$).
The $\mathbf{C}$ channel preserves the population of each eigenspace of $T$ and allows the dephasing strengths to differ. Within $\mathcal V_\pm$, it applies $P$ with probability $q_\pm$ and the identity with probability $p_\pm=1-q_\pm$. Its block action is
\begin{equation}
\mathcal N_{\mathbf{C}}(\rho)=e^{sG_{\mathbf{C}}(\theta)+\lambda \mathbf{S}_T}(\rho)=
\begin{pmatrix}
(p_+\mathcal I+q_+\mathcal C_P)(\rho_{++})
 & \eta(c_0\mathcal I+c_1\mathcal C_P)(\rho_{+-})\\[2pt]
\eta(c_0\mathcal I+c_1\mathcal C_P)(\rho_{-+})
 & (p_-\mathcal I+q_-\mathcal C_P)(\rho_{--})
\end{pmatrix}.
\label{eq:Csectorblocks}
\end{equation}
The real coefficients $c_0$ and $c_1$ weight the two contributions to coherence between eigenspaces. The conjugations by $P$ and $Q$ coincide within either eigenspace, so either Pauli gives the same dephasing there.

For $s>0$, dephasing is stronger in $\mathcal V_+$ when $\sin 2\theta$ is positive and in $\mathcal V_-$ when it is negative. The strengths coincide at $\sin 2\theta=0$. For example, take $P=Z_2$, $Q=Z_1Z_2$ and $\theta=-\pi/4$, so that $T=Z_1$. A computational-basis measurement of qubit 1 identifies the action on qubit 2. Outcome 1 selects $Z_2$ dephasing and outcome 0 selects the identity.

\emph{Population transfer between subspaces}~($\mathbf{A}$).
The $\mathbf{A}$ channel transfers population between the eigenspaces of $T$. An input in $\mathcal V_\pm$ moves to the opposite eigenspace under $P$ with probability $q_\pm$ and stays in its initial eigenspace with probability $p_\pm=1-q_\pm$. For a general input, the block action is
\begin{equation}
\mathcal N_{\mathbf{A}}(\rho)=e^{sG_{\mathbf{A}}(\theta)+\lambda \mathbf{S}_T}(\rho)=
\begin{pmatrix}
p_+\rho_{++}+q_-P\rho_{--}P
 & \eta(c_0\rho_{+-}+c_1P\rho_{-+}P)\\[2pt]
\eta(c_0\rho_{-+}+c_1P\rho_{+-}P)
 & p_-\rho_{--}+q_+P\rho_{++}P
\end{pmatrix}.
\label{eq:Asectorblocks}
\end{equation}
Each diagonal block combines the contribution that stays in its eigenspace with the contribution transferred from the opposite eigenspace. Conjugation by $P$ can also couple the two off-diagonal coherence blocks. For $s>0$, transfer favors $\mathcal V_-$ when $\sin 2\theta>0$ and $\mathcal V_+$ when $\sin 2\theta<0$. To recover amplitude damping exactly, set $P=X$, $Q=Y$, $T=Z$, $\theta=-\pi/4$ and $\lambda=0$. The channel fixes $|0\rangle$, sends $|1\rangle$ to $|0\rangle$ with probability $1-e^{-2s}$, and multiplies $\rho_{01}$ by $e^{-s}$. More generally, the additional dephasing factor $\eta$ acts only on the off-diagonal blocks and leaves the transfer probabilities fixed.

\emph{Event-correlated phase kicks}~($\mathbf{E}$).
In the $\mathbf{E}$ family, one event applies a Pauli operation and a phase rotation together. For commuting $P,Q$, the unitary for one event is
\begin{equation}
\label{eq:rotatedEgenerator}
U_\theta=\cos\theta\,P-i\sin\theta\,Q.
\end{equation}
The channel draws the number of events from a Poisson distribution with mean $s$ and applies the corresponding power of $U_\theta$, giving
\begin{equation}
\mathcal N_{\mathbf{E}}(s,\theta)=e^{sG_{\mathbf{E}}(\theta)}
=\sum_{m=0}^{\infty}\frac{e^{-s}s^m}{m!}\,\mathcal C_{U_\theta^m}.
\label{eq:Ephysicalkraus}
\end{equation}

For example, take $P=X\otimes I$ and $Q=X\otimes Z$. Each application of $U_\theta=X\otimes e^{-i\theta Z}$ flips qubit 1 and rotates qubit 2. In $U_\theta^m$, the parity of $m$ determines whether qubit 1 flips overall, and $m\theta$ determines the rotation of qubit 2.

\emph{Rotated-axis dephasing}~($\mathbf{F}$).
For anticommuting $P,Q$, the angle $\theta$ selects the dephasing axis of the $\mathbf{F}$ channel. The corresponding rotated Pauli operator is
\begin{equation}
\label{eq:rotatedFgenerator}
V_\theta=\cos\theta\,P+\sin\theta\,Q.
\end{equation}
As in Pauli dephasing, the channel is a mixture of the identity and conjugation by the selected operator, giving
\begin{equation}
\label{eq:rotatedFchannel}
\mathcal N_{\mathbf{F}}(s,\theta)
=e^{sG_{\mathbf{F}}(\theta)}
=\frac{1+e^{-2s}}2\,\mathcal I
+\frac{1-e^{-2s}}2\,\mathcal C_{V_\theta}.
\end{equation}
The channel preserves the populations in the eigenspaces of $V_\theta$ and attenuates the coherence between them by a factor $e^{-2s}$. For $P=Z$, $Q=X$ and $\theta=\pi/4$, the rotated operator is the Hadamard gate, $V_{\pi/4}=(X+Z)/\sqrt2=H_{\rm gate}$. The channel applies $H_{\rm gate}$ with probability $(1-e^{-2s})/2$ and the identity otherwise. Both probabilities approach $1/2$ as $s\to\infty$.


\section{Beyond Pauli noise}
\label{sec:qec-models}

Random Pauli operations provide a simple model of circuit noise. We use our channel families to model a broader range of effects. The following examples show how several noise mechanisms can act together on a single qubit, how population transfer biases joint parity, and how leakage occupation affects subsequent operations.

\subsection{Single-qubit noise}
\label{sec:damping-example}

Rotation, dephasing and population transfer can act together on a single qubit. The $\mathbf{H}$ family gives coherent rotations, $\mathbf{S}$ and $\mathbf{F}$ give dephasing, and $\mathbf{A}$ gives relaxation or excitation.
For a single-qubit noise channel $\Phi$, let $p=\tfrac12\|\Phi-\mathcal I\|_\diamond$ be its noise strength. For $p<\tfrac12$, allowing $s$ to take negative values lets a finite product reproduce $\Phi$ exactly. With all strengths $s\geq0$, a finite product of physical channels approximates $\Phi$ with diamond-norm error $O(p^2)$ (see the Supplemental Material).

To combine relaxation toward $|0\rangle$ with dephasing and phase rotation about $Z$, choose $P=X$ and $Q=Y$. The exact composite channel is
\begin{equation}
\mathcal N_{\mathbf{H}}(\theta)\circ\mathcal N_{\mathbf{S}}(\lambda)\circ
\mathcal N_{\mathbf{A}}(s/2,-\pi/4,0),
\qquad s,\lambda\ge0,\quad\theta\in\mathbb R.
\label{eq:qubit-composed-noise}
\end{equation}
The excited-state population retains a fraction $e^{-s}$ of its initial value, and $\rho_{+-}$ is multiplied by $e^{-s/2-4\lambda-2i\theta}$. The relaxation strength $s$ affects both quantities. The parameters $\lambda$ and $\theta$ independently control the additional decay and the phase shift, respectively.

\subsection{Non-unital noise in joint parity}
\label{sec:bias}

A channel is \emph{unital} if it preserves the maximally mixed state $I/2^n$, as every random Pauli channel does. Of the six generator classes, only $\mathbf{A}$ can act nontrivially on the identity. Contributions from several $\mathbf{A}$ terms can cancel.

Single-qubit amplitude damping changes the populations of a maximally mixed input. To see how population transfer can affect a joint state and preserve both maximally mixed marginals, take $P=X_1$ and $Q=Y_1Z_2$. Then $T=Z_1Z_2$ distinguishes even from odd $Z$ parity, and the $\mathbf{A}$ channel transfers population between the two parity eigenspaces. For $\lambda=0$ and a maximally mixed input, this gives
\begin{equation}
\mathcal N_{\mathbf{A}}(s,\theta,0)(I/4)
=\tfrac14\bigl[I-(1-e^{-2s})\sin 2\theta\, Z_1Z_2\bigr].
\label{eq:two-qubit-parity-bias}
\end{equation}
For $s>0$ and $\sin 2\theta\ne0$, a joint parity measurement detects the population imbalance. Each qubit still has reduced state $I/2$, so neither marginal reveals the bias.

\subsection{Leakage model and correlated disturbances}
\label{sec:leakage}
\label{sec:conditional-example}
\label{sec:flag-records}

The effect of a circuit operation can depend on whether a system occupies a leakage subspace~\cite{wood2018,suchara2015}. We model this dependence with an effective four-level system consisting of a two-dimensional computational subspace $\mathcal H_{\rm C}$ and a two-dimensional leakage subspace $\mathcal H_{\rm L}$. The internal coordinate $d$ indexes the basis states $j=0,1$ within each subspace. The \emph{leakage register} $v$ identifies the subspace, with $v=0$ for a computational state and $v=1$ for a leaked state. This choice gives the representation
\begin{equation}
\begin{aligned}
\mathcal H_{\rm C}\oplus\mathcal H_{\rm L}&\simeq\mathbb C_d^2\otimes\mathbb C_v^2,\\
|j\rangle_{\rm C}&\mapsto|j\rangle_d|0\rangle_v,\qquad
|j\rangle_{\rm L}\mapsto|j\rangle_d|1\rangle_v,\quad j=0,1.
\end{aligned}
\label{eq:leakage-factorization}
\end{equation}
The coordinates $d$ and $v$ describe the original four levels. Related QEC simulations use two-qubit encodings of qutrit leakage models~\cite{camps2024}. As in coherent-leakage QEC models~\cite{manabe2025}, the full density operator can retain coherence between the computational and leakage subspaces. For inputs that are \emph{block diagonal} in these subspaces and channels that preserve this property, a classical indicator can track the occupation~\cite{suchara2015,marshall2025}. A computational input has $v=0$, and several circuit operations can share the same register.

A computational-basis measurement of $v$ supplies the decoder with a classical report of the occupied subspace. A leakage report can trigger a reset to the fresh computational state $|0\rangle_d|0\rangle_v$. State-independent one-way leakage gives a known-location erasure when the encoded information in the leakage subspace is completely lost and $v$ is read out perfectly~\cite{wu2022}.

We use three families to describe population transfer, dephasing during leakage, and rotations of a neighbor that accompany a change of subspace.

\emph{Leakage and return}~($\mathbf{A}$). Leakage transfers population from $\mathcal H_{\rm C}$ to $\mathcal H_{\rm L}$, and return transfers it in the reverse direction. To control the two transition strengths independently, choose
\begin{equation}
P=I_d\otimes X_v,\qquad Q=I_d\otimes Y_v,\qquad T=I_d\otimes Z_v.
\label{eq:leakage-A-axes}
\end{equation}
The leakage channel $\mathcal N_{\mathbf{A}}(\ell/2,\pi/4,0)$ transfers an input from $\mathcal H_{\rm C}$ to $\mathcal H_{\rm L}$ with probability $1-e^{-\ell}$. The return channel $\mathcal N_{\mathbf{A}}(r/2,-\pi/4,0)$ transfers an input from $\mathcal H_{\rm L}$ back with probability $1-e^{-r}$. Here $\ell,r\ge0$. Both channels act only on $v$, carrying the input state of $d$ between the paired subspaces. One-way leakage makes that subspace absorbing. However, if it returns, the return transfers the internal state with whatever disturbance in the leaked subspace back to the computational subspace.

\emph{Noise conditional on leaked occupation}~($\mathbf{C}$). Let $R$ be a nonidentity Pauli on $d$ or on a neighboring qubit. To dephase along $R$ when the system occupies the leakage subspace, choose
\begin{equation}
P=R\otimes I_v,\qquad Q=R\otimes Z_v,\qquad T=I\otimes Z_v.
\label{eq:leakage-axes}
\end{equation}
The channel $\mathcal N_{\mathbf{C}}(s/2,-\pi/4,0)$ preserves both subspace populations. It leaves the computational block intact and dephases the affected qubit along $R$ in the leakage block. Because the dephasing depends on $v$, the affected qubit can become correlated with the leakage occupation.

Within the leakage block, dephasing along $X_d$ preserves the populations in the $X_d$ basis, and dephasing along $Z_d$ preserves those in the $Z_d$ basis. In the composition of the two channels, taking both strengths to infinity makes $d$ maximally mixed and removes its correlations with other systems conditional on leakage. The encoded information is then completely lost, as required in the erasure limit.

\emph{Neighbor rotation at a transition}~($\mathbf{E}$). A transition can itself disturb a neighboring qubit. Partner phases in transmon controlled-$Z$ (CZ) gates motivate this example~\cite{varbanov2020}. For a neighbor $a$ and a Pauli $R_a\ne I_a$, choose
\begin{equation}
P=I_a\otimes X_v,\qquad Q=R_a\otimes X_v,\qquad
U_\theta=e^{-i\theta R_a}\otimes X_v.
\label{eq:neighbor-pulse}
\end{equation}
Here $T=R_a\otimes I_v$. Each event flips $v$ and rotates the neighbor. The occupied subspace changes for odd $m$ and stays the same for even $m$. Both directions therefore have the same switching probability. For either parity, the rotation of the neighbor after $m$ events is $e^{-im\theta R_a}$.

If two channels depend on the same unobserved leakage occupation, errors at their circuit locations can be correlated. Ideal readout removes coherence between the subspaces. Discarding its report preserves the distribution of $v$ and can therefore preserve the correlations. A disturbance on the neighbor can also survive readout.


\section{Discussion}
\label{sec:discussion}

In this article, we have constructed six families of physical channels from elementary error generators in the Pauli basis. Their composition provides a systematic way to model non-Pauli errors such as coherent rotation accompanied by relaxation and dephasing. When successive gates interact with a shared leakage register, the same construction can capture non-Markovian noise on the data qubits. Leakage-induced correlations across rounds have been observed in superconducting QEC experiments~\cite{mcewen2021}.

Our framework uses compact channel models to predict measurable differences between noise processes. For fixed Pauli axes, at most three continuous parameters per channel determine how populations and coherences change. The family that reproduces single-qubit amplitude damping can also bias the parity of a maximally mixed qubit pair while preserving both marginals. Several leakage models prepare a fixed qubit state on return~\cite{suchara2015,wood2018,battistel2021}. In our model, dephasing in the two-dimensional leakage subspace controls how much coherence survives return for fixed observed leakage and return probabilities.

These channel models provide a starting point for connecting measurements of physical noise to circuit-level predictions. One direction is to identify the noise mechanisms relevant to an experimental system, use our framework to build a model for that system, and fit its parameters to the measured data. Experimental fits of sparse Pauli--Lindblad models provide a precedent~\cite{vandenberg2023}. A second direction is to use the generator algebra to reduce the computational cost of simulating QEC circuits with non-Pauli noise~\cite{taxonomy2022,hines2026}. Which channel sequences can we combine before simulation, and how much does this reduce runtime for a given accuracy in the predicted logical error rate?

\section*{AI Disclosure}

This manuscript was written in collaboration with GPT.

\section*{Acknowledgments}

We thank Mile Gu, Jayne Thompson, and Hailan Ma for their helpful suggestions and discussions. We also thank Binbin Li for the excellent wine that sparked new ideas.

%

\appendix
\section*{Supplementary Theory for Pauli-Axis Noise Models}

We define the Pauli-axis generator basis and derive the block formulas and Kraus operators for the six channel families. We then relate the families to complete Pauli jumps, prove the single-qubit composition and joint-parity results, and derive the channels and instruments for the leakage model.

\section{Pauli eigenspaces and channels}\label{supp:sec:foundations}

We introduce two tools for describing the noise channels below. The eigenspaces of a Pauli operator let us track how a channel changes populations and coherences. Choi's criterion and the Kraus representation let us verify complete positivity and trace preservation.

\subsection{Pauli axes and subspaces}

The Pauli basis provides coordinates for the generators. The eigenspaces of a chosen Pauli axis identify the populations and coherences on which the resulting channels act.

On $n\ge1$ qubits, let $\mathcal P_n=\{I,X,Y,Z\}^{\otimes n}$ be the unsigned Pauli strings, with $I$ denoting the identity operator. These strings are Hermitian unitaries. For $P,Q\in\mathcal P_n$,
\begin{equation}
\tr(PQ)=2^n\,\delta_{P,Q}.
\label{lem:pauli}
\end{equation}
Indeed, this identity holds for the four single-qubit matrices and traces multiply under tensor products. The tensor products of the four single-qubit basis matrices give all $4^n$ members of this operator basis. The coefficients of Hermitian operators in this basis are real. A single-qubit Pauli with a qubit subscript, such as $Z_j$, acts on that qubit and as the identity on the others.

A nonidentity Pauli axis is a nonidentity Pauli string with either sign.

\begin{proposition}[Eigenspaces of a signed Pauli axis]\label{prop:pauli-sector-geometry}
Let $T$ be a nonidentity Pauli axis on $n\ge1$ qubits. The projectors $\Pi_\sigma=(I+\sigma T)/2$, $\sigma\in\{+1,-1\}$, have orthogonal ranges $\mathcal V_\sigma$ of rank $2^{n-1}$. They satisfy $T\Pi_\sigma=\sigma\Pi_\sigma$, so $T\psi=\Pi_+\psi-\Pi_-\psi$ for every state vector $\psi$.
\end{proposition}
\begin{proof}
Using $T^\dagger=T$ and $T^2=I$ gives $\Pi_\sigma^\dagger=\Pi_\sigma$, $\Pi_\sigma^2=\Pi_\sigma$, $\Pi_+\Pi_-=0$, $\Pi_++\Pi_-=I$ and $T\Pi_\sigma=\sigma\Pi_\sigma$. Since $T$ is orthogonal to $I$ in Eq.~\eqref{lem:pauli}, $\tr T=0$. Taking the trace of $\Pi_\sigma$ gives its rank.
\end{proof}

For a matrix $X$ write $X_{\sigma\tau}=\Pi_\sigma X\Pi_\tau$. For a density operator $\rho\ge0$ with $\tr\rho=1$, the Born rule gives $\tr(\Pi_\sigma\rho)=\tr\rho_{\sigma\sigma}$ as the probability of outcome $\sigma$. The off-diagonal blocks $\rho_{\sigma,-\sigma}$ carry coherence between the subspaces.

\subsection{Channels and Kraus operators}\label{sec:channelfoundations}

The eigenspace blocks describe the input to a channel. We use the Choi matrix and the Kraus representation to check that the proposed maps remain physical on inputs entangled with a reference.

A superoperator is a complex-linear map on matrices. The identity superoperator is $\id(X)=X$. A channel is a superoperator that is completely positive (CP) and trace preserving (TP), together abbreviated CPTP. Complete positivity means that the map remains positive when tensored with the identity map on any reference system. A map $\Gamma$ is trace preserving if $\tr\Gamma(X)=\tr X$ for all $X$. In particular, $\id$ is a channel.

The Choi matrix of a superoperator $\Gamma$ is
$J(\Gamma)=\sum_{j,k}\Gamma(|j\rangle\langle k|)\otimes |j\rangle\langle k|$
in a fixed orthonormal basis. Write $|L\rangle\rangle=\sum_j L|j\rangle\otimes|j\rangle$. The map $X\mapsto LXM^\dagger$ has Choi matrix $|L\rangle\rangle\langle\langle M|$ by expansion. The matrix units form a basis, so $J$ is injective. A Hermiticity-preserving map sends adjoints to adjoints by complex linearity, and therefore has a Hermitian Choi matrix.

\begin{theorem}[Choi's criterion and Kraus representation~{\cite{choi1975}}]\label{lem:choi}
A complex-linear map $\Gamma$ on a finite-dimensional matrix space is CP if and only if $J(\Gamma)\ge0$. This is equivalent to a finite representation
\begin{equation}
\Gamma(X)=\sum_k K_k X K_k^\dagger.
\label{eq:kraus-representation}
\end{equation}
For such a representation, $\Gamma$ is TP if and only if $\sum_k K_k^\dagger K_k=I$.
\end{theorem}
\begin{proof}
If $\Gamma$ is CP, its extension by the identity sends $|I\rangle\rangle\langle\langle I|$ to $J(\Gamma)\ge0$. If $J(\Gamma)$ is positive, factor it into outer products of vectors $|K_k\rangle\rangle$. Injectivity of $J$ gives Eq.~\eqref{eq:kraus-representation}. Each sandwich is positive on any reference because its extended Kraus matrix is $K_k\otimes I$. Finally, trace cyclicity gives $\tr(\Gamma(X))=\tr(X\sum_k K_k^\dagger K_k)$ for every $X$. Nondegeneracy of the trace pairing proves the TP criterion.
\end{proof}

Compositions and probability mixtures of channels are CPTP. Norm limits of channels are CPTP because positivity of the Choi matrix and trace preservation are closed conditions. This also applies to a convergent Kraus sum with $\sum_k K_k^\dagger K_k=I$. The Choi matrices of its partial sums are positive, and taking the trace of the limit gives trace preservation.

\section{Generators and physical completions}\label{sec:algebra}

We construct the generators of the six channel families. Starting from a Pauli basis for trace-annihilating, Hermiticity-preserving maps, we group its elements by their axes and relative phases. For each two-axis class, we specify the dephasing terms that complete the generator and the parameter range used when taking its exponential.

\subsection{Exponentials and the Pauli-axis basis}

A trace-annihilating, Hermiticity-preserving generator gives an exponential that preserves both trace and Hermiticity. A Pauli-coordinate basis for these generators lets us study complete positivity through explicit families.

For a superoperator $G$, define $e^G=\sum_{k\ge0}G^k/k!$, with powers given by composition and $G^0=\id$. Absolute convergence follows from submultiplicativity in an induced norm $\|\cdot\|$. A map $G$ is trace annihilating if $\tr G(X)=0$ for every $X$. Together with Hermiticity preservation, this defines the real vector space of trace-annihilating and Hermiticity-preserving (TAHP) maps.

\begin{proposition}[Exponentials of TAHP maps]\label{prop:tahp-exponential}
If $G$ is TAHP, then $e^G$ preserves trace and Hermiticity and is a channel exactly when it is CP. In an induced norm, $\|e^G-\id\|\le e^{\|G\|}-1$, so $e^G\to\id$ as $\|G\|\to0$.
\end{proposition}
\begin{proof}
For $k\ge1$, $\tr G^k(X)=0$ because $G^k(X)=G(G^{k-1}(X))$. Only $G^0$ contributes to the trace of the exponential. Each power preserves Hermiticity, as does the convergent real series. Bounding each term of $e^G-\id$ by $\|G\|^k/k!$ gives the norm bound.
\end{proof}

For matrices $B,D$, use the commutator $[B,D]=BD-DB$ and anticommutator $\{B,D\}=BD+DB$. For $P,Q\in\mathcal P_n$ and $\alpha=0,1$, define
\begin{equation}
\mathcal F_{PQ}(\alpha)(X)=i^\alpha PXQ+i^{-\alpha}QXP
-\tfrac12\{i^\alpha QP+i^{-\alpha}PQ,X\}.
\label{prop:F_properties}
\end{equation}
The same formula applies to signed Pauli strings. The two sandwich terms are adjoints for Hermitian $X$ and the matrix in the anticommutator is Hermitian. The cyclic trace of the two sandwiches is $\tr[(i^\alpha QP+i^{-\alpha}PQ)X]$, which is canceled by the anticommutator. Thus every $\mathcal F_{PQ}(\alpha)$ is TAHP.

\begin{theorem}[Pauli-axis generator basis in EEG coordinates~{\cite{taxonomy2022,miller2025}}]\label{lem:generatorbasis}
Fix a total order $<$ on the nonidentity elements of $\mathcal P_n$. The maps $\mathcal F_{IP}(1)$ and $\mathcal F_{PP}(0)$ for $P\ne I$, together with $\mathcal F_{PQ}(0)$ and $\mathcal F_{PQ}(1)$ for distinct nonidentity $P<Q$, form a real basis of TAHP maps.
\end{theorem}
\begin{proof}
The Pauli vectors $|R\rangle\rangle$ give an orthogonal Choi basis. Write the Choi matrix of a TAHP map in outer products of these vectors, and let $(C_{RS})_{R,S\ne I}$ be its nonidentity coefficient block. This matrix is Hermitian. Subtract the TAHP map
\[
X\longmapsto\sum_{R,S\ne I} C_{RS}\bigl(RXS-\tfrac12\{SR,X\}\bigr).
\]
The anticommutator terms involve left or right multiplication alone, so only the sandwich terms contribute to the nonidentity block. Their coefficients are exactly $C_{RS}$. The subtracted map is TAHP by trace cyclicity and $C_{SR}=C_{RS}^*$. Hermiticity of the remaining Choi matrix makes the difference $aX+BX+XB^\dagger$, where $a$ is real and $B$ is traceless. Trace annihilation implies $aI+B+B^\dagger=0$. Taking trace gives $a=0$, so $B=-iH$ for a traceless Hermitian $H$. Thus the remaining map is $-i[H,X]$.

The diagonal, real off-diagonal and imaginary off-diagonal coordinates of $(C_{RS})$ are supplied respectively by $\mathcal F_{PP}(0)$, $\mathcal F_{PQ}(0)$ and $\mathcal F_{PQ}(1)$, with a factor of two on the diagonal. The unique Pauli expansion of $H$ supplies the remaining coordinates through $\mathcal F_{IP}(1)(X)=-i[P,X]$. Uniqueness of both parts gives linear independence.
\end{proof}

\subsection{Six classes and their parameter domains}

The basis separates the single-axis terms from the cross terms. Commutation and the phase of a cross term distinguish the four two-axis classes, whose complete generators include dephasing along the chosen axes.

Table~\ref{tab:classes} defines the six classes and lists their complete generators.

For distinct nonidentity $P,Q$, choose the signed product axis
\begin{equation}
T=
\begin{cases}
PQ,& PQ=QP,\\
-iPQ,& PQ=-QP.
\end{cases}
\label{lem:pauli_multi}
\end{equation}
The single-qubit Pauli products give one unsigned string with phase in $\{1,-1,i,-i\}$. Since $(PQ)^\dagger=QP$, this phase is real for commuting pairs and imaginary for anticommuting pairs. Thus $T$ is a Hermitian nonidentity Pauli string with its sign retained. Its eigenvalues label the same subspaces throughout. Each of $P,Q$ commutes with $T$ in the first case and anticommutes with $T$ in the second. For $\mathbf{H}$ and $\mathbf{S}$, $T$ is the chosen single axis. The same defining formula for $\mathcal F$ applies to signed axes in $\mathbf{H}_T$ and $\mathbf{S}_T$. The dephasing direction $\mathbf{S}_T$ is independent of the sign of $T$, since $\mathbf{S}_T(X)=2(TXT-X)$.

For the cross-term classes $\nu$ in the table and a real mixing angle $\theta$, define
\begin{equation}
G_\nu(\theta)=\tfrac12\bigl(\cos^2\theta\,\mathbf{S}_P+\sin^2\theta\,\mathbf{S}_Q+\sin2\theta\,\nu_{P,Q}\bigr),
\qquad \nu\in\{\mathbf{C},\mathbf{A},\mathbf{E},\mathbf{F}\}.
\label{eq:angleACgenerators}
\end{equation}
The $\mathbf{H}$ rotation angle is also real. In the complete generators, $s\ge0$ sets the two-axis noise strength and $\lambda\ge0$ sets the $T$-axis dephasing strength. Each channel is the exponential of the complete generator in its row. For fixed axes each row uses at most three continuous parameters.

\begin{table}[htbp]
\centering
\caption{The six basis directions and their completed generators. In the cross-term rows the axes are distinct and nonidentity.}
\label{tab:classes}
\label{def:pauliaxisclasses}
\begin{tabular}{lll}
\toprule
Direction & Condition & Complete generator\\
\midrule
$\mathbf{H}_T=\mathcal F_{IT}(1)$ & $T\ne\pm I$ & $\theta \mathbf{H}_T$\\
$\mathbf{S}_T=\mathcal F_{TT}(0)$ & $T\ne\pm I$ & $\lambda \mathbf{S}_T$\\
$\mathbf{C}_{P,Q}=\mathcal F_{PQ}(0)$ & $PQ=QP$ & $sG_{\mathbf{C}}(\theta)+\lambda \mathbf{S}_T$\\
$\mathbf{A}_{P,Q}=\mathcal F_{PQ}(1)$ & $PQ=-QP$ & $sG_{\mathbf{A}}(\theta)+\lambda \mathbf{S}_T$\\
$\mathbf{E}_{P,Q}=\mathcal F_{PQ}(1)$ & $PQ=QP$ & $sG_{\mathbf{E}}(\theta)$\\
$\mathbf{F}_{P,Q}=\mathcal F_{PQ}(0)$ & $PQ=-QP$ & $sG_{\mathbf{F}}(\theta)$\\
\bottomrule
\end{tabular}
\end{table}

For a unitary $U$, write $\C_U(X)=UXU^\dagger$.
Direct substitution gives $\mathbf{S}_T=2(\C_T-\id)$, $\mathbf{H}_T=-i[T,\cdot]$ and $\mathcal F_{II}(\alpha)=0$. Exchanging axes leaves $\mathcal F_{PQ}(0)$ unchanged and reverses $\mathcal F_{PQ}(1)$, and $\mathcal F_{IP}(0)=\mathcal F_{PP}(1)=0$. Thus these six directions, with the fixed pair order in the basis, cover all the nonzero cases. A signed Hermitian Pauli used as an $\mathbf{H}$ axis can be implemented by reversing the angle on its unsigned axis.

A \emph{jump operator} $L$ describes an error mechanism through the map
$X\mapsto LXL^\dagger$, which has the same form as one term in a Kraus sum.
For example, $L=\lvert0\rangle\!\langle1\rvert$ describes decay from
$\lvert1\rangle$ to $\lvert0\rangle$.
The corresponding term in the generator is the dissipator
$\mathcal D_L(X)=LXL^\dagger-\{L^\dagger L,X\}/2$.
The anticommutator term cancels the trace of the first term.

For $a=\cos\theta$ and $b=\sin\theta$, direct expansion gives
\begin{equation}
\mathcal D_{aP+bQ}=a^2\mathcal D_P+b^2\mathcal D_Q+ab\mathcal F_{PQ}(0),
\qquad
\mathcal D_{aP-ibQ}=a^2\mathcal D_P+b^2\mathcal D_Q+ab\mathcal F_{PQ}(1).
\label{thm:families}
\end{equation}
As $\mathbf{S}_P=2\mathcal D_P$, the first identity gives $G_{\mathbf{C}}$ and $G_{\mathbf{F}}$ for their respective axis pairs, and the second gives $G_{\mathbf{A}}$ and $G_{\mathbf{E}}$. These identities are algebraic tools for evaluating the exponential channels.

\section{Channel actions and Kraus representations}\label{sec:finitefamilies}

We evaluate the generator exponentials and prove that all six families give physical channels throughout their stated parameter ranges. We begin with single-axis rotation and dephasing, then derive the block maps and Kraus operators for subspace-dependent dephasing and population transfer. For the final two families, we express the channel as a mixture of unitary operations to identify the phase kicks and the rotated dephasing axis.

\subsection{Pauli rotations and dephasing (\texorpdfstring{$\mathbf{H}$ and $\mathbf{S}$}{H and S})}

With a single Pauli axis, the $\mathbf{H}$ and $\mathbf{S}$ channels act directly on the coherence between its eigenspaces. Products of the same Pauli rotations can also approximate general unitary conjugations.

\begin{proposition}[Single-axis rotation and dephasing]\label{lem:single}
Let $T$ be a nonidentity Pauli axis. For $\theta\in\mathbb R$, $\N_{\mathbf{H}}(\theta)=e^{\theta \mathbf{H}_T}=\C_{e^{-i\theta T}}$, with single Kraus operator $e^{-i\theta T}=\cos\theta\,I-i\sin\theta\,T$. Its action on the subspace blocks is
\begin{equation}
(\N_{\mathbf{H}}(\theta)(X))_{\sigma\tau}=e^{-i\theta(\sigma-\tau)}X_{\sigma\tau}.
\label{eq:singleHblocks}
\end{equation}
For $\lambda\ge0$ and $\eta=e^{-4\lambda}$, the $\mathbf{S}$ channel and its Kraus list are
\begin{equation}
\N_{\mathbf{S}}(\lambda)=e^{\lambda \mathbf{S}_T}
=\tfrac{1+\eta}{2}\id+\tfrac{1-\eta}{2}\C_T,\qquad
K_0=\sqrt{\tfrac{1+\eta}{2}}\,I,\quad K_1=\sqrt{\tfrac{1-\eta}{2}}\,T.
\label{supp:eq:pauliexp}
\end{equation}
Its block action is
\begin{equation}
(\N_{\mathbf{S}}(\lambda)(X))_{\sigma\tau}=
\begin{cases}
X_{\sigma\tau},&\sigma=\tau,\\
\eta X_{\sigma\tau},&\sigma\ne\tau.
\end{cases}
\label{eq:singleSblocks}
\end{equation}
\end{proposition}
\begin{proof}
Left and right multiplication by $T$ commute. Factoring the exponential of $\theta \mathbf{H}_T$ therefore gives left multiplication by $e^{-i\theta T}$ and right multiplication by $e^{i\theta T}$. Expanding these matrices on the two projectors gives the $\mathbf{H}$ blocks. The single Kraus operator is unitary. For $\mathbf{S}$, use $\C_T^2=\id$ to expand $e^{2\lambda(\C_T-\id)}$ into even and odd powers of $\C_T$. Their weights are nonnegative and sum to one, proving completeness. On $X_{\sigma\tau}$, $\C_T$ acts by $\sigma\tau$. Substitution gives the stated $\mathbf{S}$ blocks.
\end{proof}

\begin{theorem}[Trotter product formula for bounded maps~{\cite{trotter1959}}]\label{thm:trotter-product}
Let $B_1,\ldots,B_m$ be linear maps on the same finite-dimensional space, and write $M=\sum_j\|B_j\|$ for an induced norm. For any positive integer $N$ and a fixed order of the product,
\begin{equation}
\left\|e^{\sum_j B_j}-\left(\prod_j e^{B_j/N}\right)^N\right\|
\le \frac{M^2e^M}{N}.
\label{eq:product-convergence}
\end{equation}
In particular the product converges to $e^{\sum_j B_j}$ in norm as $N\to\infty$.
\end{theorem}
\begin{proof}
The series for $e^{\sum_j B_j/N}$ and $\prod_j e^{B_j/N}$ have the same constant and linear terms. Submultiplicativity bounds the remaining terms of either series by $e^{M/N}-1-M/N\le (M/N)^2e^{M/N}/2$. Each map has norm at most $e^{M/N}$. The telescoping expansion of the difference of the $N$th powers has $N$ terms, each bounded by $e^{(N-1)M/N}$ times the single-factor difference. This gives Eq.~\eqref{eq:product-convergence}.
\end{proof}

\begin{proposition}[Approximation by Pauli-axis rotations]\label{cor:paulitrotter}
For any unitary $U$ on $n$ qubits, products of $\mathbf{H}$ rotations about Pauli axes approximate $\C_U$ to arbitrary accuracy in an induced norm.
\end{proposition}
\begin{proof}
For a unitary $U$, its spectral decomposition gives a Hermitian operator $H_U$ with $U=e^{-iH_U}$. Expand $H_U$ in the Pauli basis. Its identity term contributes only a global phase. For the real coefficients $h_P$ of the other terms, apply Eq.~\eqref{eq:product-convergence} to the maps $h_P \mathbf{H}_P$. This gives
\begin{equation}
\C_U=\lim_{N\to\infty}\left(\prod_{P\ne I}e^{(h_P/N)\mathbf{H}_P}\right)^N ,
\label{eq:paulitrotter}
\end{equation}
with a fixed order for the finite product.
\end{proof}

\subsection{Subspace-dependent dephasing and population transfer (\texorpdfstring{$\mathbf{C}$ and $\mathbf{A}$}{C and A})}

With two axes, the branch weights can depend on the input eigenspace. A common Kraus construction captures dephasing within eigenspaces for $\mathbf{C}$ and population transfer between them for $\mathbf{A}$.

Let $\Pi_\pm$ be complementary orthogonal projectors and $P$ be unitary. In branch $j\in\{0,1\}$, write $w_{j,\pm}$ for the weights on the diagonal blocks and $g_j,g_j^*$ for the multipliers on the $+-$ and $-+$ blocks. Each branch conjugates by $P^j$ after applying these multipliers. The weights and multipliers satisfy
\begin{equation}
w_{j,\pm}\ge0,\qquad
w_{0,\pm}+w_{1,\pm}=1,\qquad
|g_j|^2\le w_{j,+}w_{j,-},
\label{eq:sectorpairaction}
\end{equation}
For $w_{j,+}>0$, the Kraus operators of exponential maps for generators $\mathbf{A}/\mathbf{C}$ that give these block multipliers are
\begin{align}
K_{j,0}&=P^j\left(\sqrt{w_{j,+}}\,\Pi_+
                  +\frac{g_j^*}{\sqrt{w_{j,+}}}\,\Pi_-\right),\notag\\
K_{j,1}&=P^j\sqrt{w_{j,-}-\frac{|g_j|^2}{w_{j,+}}}\,\Pi_- .
\label{eq:sectorweightkraus}
\end{align}
When $w_{j,+}=0$, Eq.~\eqref{eq:sectorpairaction} forces $g_j=0$, and we take $(K_{j,0},K_{j,1})=(\sqrt{w_{j,-}}\,P^j\Pi_-,0)$.
The Kraus form gives complete positivity. Since $\sum_k K_{j,k}^\dagger K_{j,k}=w_{j,+}\Pi_++w_{j,-}\Pi_-$, branch $j$ has probability $w_{j,\sigma}$ for a normalized input in $\mathcal V_\sigma$, and the weight normalization makes the sum of the branches trace preserving.

\begin{theorem}[$\mathbf{C}$/$\mathbf{A}$ block maps and Kraus operators]
\label{cor:angleACfamilies}
\label{cor:canonicalKsectors}
\label{prop:C-dephasing-blocks}
Choose distinct nonidentity Pauli strings $P,Q$, commuting for $\mathbf{C}$ and anticommuting for $\mathbf{A}$, and let $T$ be as in Eq.~\eqref{lem:pauli_multi}. For $s,\lambda\ge0$ and $\theta\in\mathbb R$, put $\eta=e^{-4\lambda}$. For input in $\mathcal V_\sigma$, $\sigma=\pm1$, branch $j=0$ has probability $p_\sigma=1-q_\sigma$, and the probability $q_\sigma$ of branch $j=1$ is
\begin{align}
\mathbf{C}:\quad q_\sigma&=\tfrac12(1-e^{-2s(1+\sigma\sin 2\theta)}),\notag\\
\mathbf{A}:\quad q_\sigma&=\tfrac12(1+\sigma\sin 2\theta)(1-e^{-2s}).
\label{eq:angleACweights}
\end{align}
Define $c_0,c_1$ by
\begin{equation}
c_0+c_1=e^{-2s\sin^2\theta},\qquad
c_0-c_1=e^{-2s\cos^2\theta}.
\label{eq:bareACcoherences}
\end{equation}
For $\nu=\mathbf{C},\mathbf{A}$, write $\N_\nu(s,\theta,\lambda)=e^{sG_\nu(\theta)+\lambda \mathbf{S}_T}$. With these parameters understood, the output blocks in the eigenspaces of $T$ are
\begin{align}
(\N_{\mathbf{C}}(X))_{\sigma\sigma}
 &=p_\sigma X_{\sigma\sigma}+q_\sigma P X_{\sigma\sigma}P,\notag\\
(\N_{\mathbf{C}}(X))_{\sigma,-\sigma}
 &=\eta(c_0 X_{\sigma,-\sigma}+c_1P X_{\sigma,-\sigma}P),\notag\\
(\N_{\mathbf{A}}(X))_{\sigma\sigma}
 &=p_\sigma X_{\sigma\sigma}+q_{-\sigma}P X_{-\sigma,-\sigma}P,\notag\\
(\N_{\mathbf{A}}(X))_{\sigma,-\sigma}
 &=\eta(c_0 X_{\sigma,-\sigma}+c_1P X_{-\sigma,\sigma}P).
\label{eq:canonicalKblocks}
\end{align}
Both channels have the Kraus operators in Eq.~\eqref{eq:sectorweightkraus} with $w_{0,\sigma}=p_\sigma$, $w_{1,\sigma}=q_\sigma$ and $g_j=\eta c_j$, using the zero-weight convention above. These branch probabilities require no subspace measurement.
\end{theorem}
\begin{proof}
Set $a=\cos\theta$, $b=\sin\theta$, $k=\sin2\theta$ and $d=\cos2\theta$. The jump for $\mathbf{C}$ is $aP+bQ$ and that for $\mathbf{A}$ is $aP-ibQ$, by Eq.~\eqref{thm:families}. In both cases the jump is $L=P(aI+bT)$ since $Q=PT$ for $\mathbf{C}$ and $Q=iPT$ for $\mathbf{A}$. Hence $L^\dagger L=I+kT$. A block $X_{\sigma\tau}$ contributes $(a+b\sigma)(a+b\tau)P X_{\sigma\tau}P$ through the jump sandwich, and $-[1+k(\sigma+\tau)/2]X_{\sigma\tau}$ through the anticommutator. The sandwich keeps both indices for $\mathbf{C}$ and reverses both for $\mathbf{A}$ because $P\Pi_\sigma=\Pi_\sigma P$ for $\mathbf{C}$ and $P\Pi_\sigma=\Pi_{-\sigma}P$ for $\mathbf{A}$. In both cases the diagonal and off-diagonal subspaces of operator space are invariant. The map $\mathbf{S}_T$ commutes with the generator because it vanishes on the diagonal blocks and acts as a scalar on the off-diagonal blocks.

For $\mathbf{C}$, the restriction of $G_{\mathbf{C}}$ to the diagonal block in $\mathcal V_\sigma$ is $(1+\sigma k)(\C_P-\id)$. Since $\C_P^2=\id$, exponentiating $s$ times this restriction gives $q_\sigma$ in Eq.~\eqref{eq:angleACweights}. The identity $Q\Pi_\sigma=\sigma P\Pi_\sigma$ gives equal Pauli conjugations within that subspace. For $\mathbf{A}$, identify $X_{++}$ and $P X_{--}P$ as operators on $\mathcal V_+$. On this pair of blocks, $G_{\mathbf{A}}$ is represented by
\[
M=\begin{pmatrix}-(1+k)&1-k\\ 1+k&-(1-k)\end{pmatrix}.
\]
Each column corresponds to the subspace of origin. Since $M^2=-2M$, $e^{sM}=I_2+(1-e^{-2s})M/2$. The coefficient of transfer from $\mathcal V_\sigma$ to the opposite subspace is therefore $q_\sigma$ in Eq.~\eqref{eq:angleACweights}, giving the diagonal blocks in Eq.~\eqref{eq:canonicalKblocks}.

For the off-diagonal blocks the generator is $-\id+d\C_P$. The exponential multiplies the $+1$ and $-1$ parts under $\C_P$ by $e^{-s(1-d)}$ and $e^{-s(1+d)}$, which are $c_0+c_1$ and $c_0-c_1$ in Eq.~\eqref{eq:bareACcoherences}. Multiplying by $\eta$ accounts for the independent $T$ dephasing. Whether $P$ reverses the two block indices then gives the two off-diagonal formulas in Eq.~\eqref{eq:canonicalKblocks}.

It remains to check the Kraus bounds in Eq.~\eqref{eq:sectorpairaction}. Both families have $0\le q_\sigma\le1$ since $s\ge0$ and $|k|\le1$. Let $u=e^{-s}$, so $c_0=u\cosh(sd)$ and $c_1=u\sinh(sd)$. Using Eq.~\eqref{eq:angleACweights},
\begin{align*}
\mathbf{C}:\quad q_+q_- - c_1^2
 &=u^2\bigl[\sinh^2s-\sinh^2(sk)-\sinh^2(sd)\bigr],\\
\mathbf{A}:\quad q_+q_- - c_1^2
 &=u^2\bigl[d^2\sinh^2s-\sinh^2(sd)\bigr].
\end{align*}
Both expressions are nonnegative because $k^2+d^2=1$ and $|\sinh(sx)|\le |x|\sinh s$ for $|x|\le1$, by oddness and convexity of $\sinh$ on $[0,s]$. Also $c_0^2-c_1^2=u^2$. Thus
\[
p_+p_- - c_0^2=(q_+q_- -c_1^2)+(1-q_+-q_--u^2).
\]
The last parenthesis is $u^2(\cosh(2sk)-1)$ for $\mathbf{C}$ and zero for $\mathbf{A}$. The inequalities persist after multiplying the coherence coefficients by $\eta\le1$, so the parameters of both families satisfy Eq.~\eqref{eq:sectorpairaction}, including zero weights and either sign of $c_1$.
\end{proof}

For $s>0$, Eq.~\eqref{eq:angleACweights} gives stronger $\mathbf{C}$ dephasing in $\mathcal V_+$ for $\sin 2\theta>0$ and in $\mathcal V_-$ for $\sin 2\theta<0$. The strengths agree when $\sin 2\theta=0$. For $P=Z_2$, $Q=Z_1Z_2$, the axis is $T=Z_1$. At $\theta=-\pi/4$, $q_+=0$ and $q_-=(1-e^{-4s})/2$. Hence a computational-basis check on qubit 1 selects the identity target action for outcome 0 and the indicated Z dephasing for outcome 1, evaluated on the input's corresponding unnormalized block. Without that check the cross blocks retain their stated action, with $c_0=e^{-s}$ and $c_1=0$.

For $\mathbf{A}$, $q_+-q_-=(1-e^{-2s})\sin 2\theta$. At $s>0$ its sign gives the direction with larger transfer probability, into $\mathcal V_-$ for positive sign and into $\mathcal V_+$ for negative sign. At $s=0$, both $\mathbf{C}$ and $\mathbf{A}$ give the $\mathbf{S}$ channel along $T$ because $q_\sigma=c_1=0$ and $c_0=1$.

\begin{proposition}[Single-qubit amplitude damping]\label{prop:amplitude-damping}
For $P=X$, $Q=Y$ and $T=Z$, the channel $\N_{\mathbf{A}}(s,-\pi/4,0)$ with $s\ge0$ has the Kraus operators
\begin{equation}
K_0=|0\rangle\langle 0|+e^{-s}|1\rangle\langle 1|,
\qquad K_1=\sqrt{1-e^{-2s}}\,|0\rangle\langle1|.
\label{eq:ampchannel}
\end{equation}
It fixes $|0\rangle\langle0|$, transfers population from $|1\rangle$ to $|0\rangle$ with probability $1-e^{-2s}$, and multiplies $\rho_{01}$ by $e^{-s}$. The angle $\pi/4$ exchanges the source and target.
\end{proposition}
\begin{proof}
Substituting $\theta=-\pi/4$ into Eqs.~\eqref{eq:angleACweights} and \eqref{eq:bareACcoherences} gives $p_+=1$, $p_-=e^{-2s}$, $c_0=e^{-s}$, $q_+=c_1=0$ and $q_-=1-e^{-2s}$. With $\eta=1$, Eq.~\eqref{eq:sectorweightkraus} reduces to Eq.~\eqref{eq:ampchannel}. Applying the two Kraus sandwiches gives the population and coherence factors. Reversing the angle interchanges $q_+$ and $q_-$ and leaves the coherence factors unchanged.
\end{proof}

\subsection{Correlated pulses and tilted dephasing axes (\texorpdfstring{$\mathbf{E}$ and $\mathbf{F}$}{E and F})}

The jump operators for $\mathbf{E}$ and $\mathbf{F}$ are unitary. Commuting axes give repeated phase kicks, and anticommuting axes give dephasing about the axis selected by their linear combination.

\begin{proposition}[Correlated phase kicks and rotated dephasing]\label{lem:rotatingpulses}
Choose distinct nonidentity Pauli strings $P,Q$, strength $s\ge0$ and angle $\theta\in\mathbb R$.
For commuting $P,Q$, $U_\theta=P e^{-i\theta T}=\cos\theta\,P-i\sin\theta\,Q$ is unitary and $G_{\mathbf{E}}(\theta)=\C_{U_\theta}-\id$. Its channel and event Kraus list are
\begin{align}
\N_{\mathbf{E}}(s,\theta)&=e^{sG_{\mathbf{E}}(\theta)}=\sum_{m\ge0}p_m\,\C_{U_\theta^m},&
p_m&=e^{-s}\frac{s^m}{m!},
\label{eq:Efactoredchannel}\\
K_m&=\sqrt{p_m}\,U_\theta^m,&m&=0,1,\ldots .
\label{eq:Eeventkraus}
\end{align}
The action of each pulse power in the eigenspaces of $T$ is
\begin{equation}
\C_{U_\theta^m}(X_{\sigma\tau})
=e^{-im\theta(\sigma-\tau)}P^m X_{\sigma\tau}P^m .
\label{eq:Epulseblocks}
\end{equation}
In particular $P=X\otimes I,\ Q=X\otimes Z$ give $U_\theta=X\otimes e^{-i\theta Z}$.

For anticommuting $P,Q$, $V_\theta=\cos\theta\,P+\sin\theta\,Q$ is a Hermitian unitary and $G_{\mathbf{F}}(\theta)=\C_{V_\theta}-\id$. Then
\begin{align}
\N_{\mathbf{F}}(s,\theta)&=e^{sG_{\mathbf{F}}(\theta)}=\tfrac{1+e^{-2s}}2\,\id+\tfrac{1-e^{-2s}}2\,\C_{V_\theta},
\label{supp:eq:rotatedFchannel}\\
K_0&=\sqrt{\tfrac{1+e^{-2s}}2}\,I,\qquad
K_1=\sqrt{\tfrac{1-e^{-2s}}2}\,V_\theta .
\label{eq:rotatedFfullkraus}
\end{align}
It preserves the eigenspace populations of $V_\theta$ and attenuates their mutual coherence by $e^{-2s}$. At $\theta=0$ it is Pauli noise, and at $s=0$ it is the identity.
\end{proposition}
\begin{proof}
For $\mathbf{E}$, $Q=PT$ gives the unitary factorization of $U_\theta$. For $\mathbf{F}$, $\{P,Q\}=0$ gives $V_\theta^2=I$ and $V_\theta=V_\theta^\dagger$. The jump identities in Eq.~\eqref{thm:families}, with squared jump modulus $I$ in each case, give the two generators. For $\mathbf{E}$ expand $e^{s(\C_{U_\theta}-\id)}$ using the commuting identity map. The Poisson weights sum to one and have mean $s$ by shifting the factorial sum. Each unitary conjugation has norm one in the induced Hilbert--Schmidt norm, so the Kraus series converges absolutely. Unitarity also gives $\sum_m K_m^\dagger K_m=I$. At $s=0$ only $m=0$ contributes. Commutation gives $U_\theta^m=P^m e^{-im\theta T}$, proving Eq.~\eqref{eq:Epulseblocks} and the shared dependence of the phase and Pauli parity on $m$. The tensor-product example follows by substitution.

For $\mathbf{F}$ use $\C_{V_\theta}^2=\id$ to sum even and odd powers. The resulting nonnegative weights sum to one. Conjugation by $V_\theta$ acts by $+1$ on each diagonal eigenspace block and by $-1$ on the off-diagonal ones, giving the stated attenuation. The choice $P=Z,\ Q=X,\ \theta=\pi/4$ gives the Hadamard gate, so this example approaches an equal mixture of identity and Hadamard as $s\to\infty$.
\end{proof}

\begin{theorem}[Physicality of the six families]\label{thm:physical-families}
The exponential of each complete generator in Table~\ref{tab:classes} is CPTP for $s,\lambda\ge0$ and $\theta\in\mathbb R$. For a fixed complete generator $G$ in the table, the channels $\Gamma_x=e^{xG}$, $x\ge0$, satisfy $\Gamma_0=\id$ and $\Gamma_{x+y}=\Gamma_y\circ\Gamma_x$ for $x,y\ge0$. In particular, every increment in strength is a CPTP factor on the full register.
\end{theorem}
\begin{proof}
Proposition~\ref{lem:single} supplies the complete Kraus lists for $\mathbf{H}$ and $\mathbf{S}$. Theorem~\ref{cor:angleACfamilies} gives the $\mathbf{C}$/$\mathbf{A}$ Kraus lists, including zero branch weights. Proposition~\ref{lem:rotatingpulses} gives a convergent Poisson Kraus sum for $\mathbf{E}$ and a complete two-operator list for $\mathbf{F}$. Theorem~\ref{lem:choi} gives CPTP in each case. Multiplying $G$ by $x\ge0$ rescales the $\mathbf{H}$ angle, the $\mathbf{S}$ dephasing parameter or the cross-term strength and any added dephasing parameter. In each case it gives another complete generator in the same domain. The power series for a single map gives $e^{(x+y)G}=e^{yG}e^{xG}$.
\end{proof}

\section{Complete Pauli jumps and the channel families}
\label{sec:complete-pauli-jumps}

A dissipative jump can contain several Pauli strings with relative
amplitudes and phases. If $L=\sum_\alpha v_\alpha P_\alpha$, its
dissipator has the Pauli expansion
\begin{equation}
\mathcal D_L(B)=\sum_{\alpha,\beta}v_\alpha\overline{v_\beta}
 \left(P_\alpha B P_\beta-\tfrac12\{P_\beta P_\alpha,B\}\right).
\label{eq:complete-pauli-jump-expansion}
\end{equation}
The coefficient matrix $vv^\dagger$ has rank one for a nonzero jump.
Its off-diagonal entries record the cross terms between components of
that jump. Applying the components as separate jumps yields the
diagonal entries alone.

\begin{proposition}[One- and two-Pauli jumps in the six families]
\label{prop:complete-jump-family-restriction}
Let $P,Q$ be distinct nonidentity Pauli strings, $a=\cos\theta$ and
$b=\sin\theta$. For the two-axis generators in
Eq.~\eqref{eq:angleACgenerators}, $G_\nu(\theta)=\mathcal D_{L_\nu}$,
with $\nu$ and $L_\nu$ related by
\begin{center}
\begin{tabular}{lcc}
\toprule
Axis relation & $L_\nu=aP+bQ$ & $L_\nu=aP-ibQ$\\
\midrule
$PQ=QP$ & $\mathbf C$ & $\mathbf E$\\
$PQ=-QP$ & $\mathbf F$ & $\mathbf A$\\
\bottomrule
\end{tabular}
\end{center}
For $s,\lambda\ge0$, the $\mathbf E$ and $\mathbf F$ channels are
$e^{s\mathcal D_{L_\nu}}$. The $\mathbf A$ and $\mathbf C$ channels
add the independent single-Pauli dissipator
$\lambda\mathbf S_T=\mathcal D_{\sqrt{2\lambda}\,T}$ in the exponent.
The $\mathbf S$ generator itself is a single-Pauli dissipator,
$\mathbf S_P=2\mathcal D_P$.
\end{proposition}
\begin{proof}
The diagonal terms in Eq.~\eqref{eq:complete-pauli-jump-expansion}
give $(a^2\mathbf S_P+b^2\mathbf S_Q)/2$. The relative phase $0$
gives the real cross term $ab\,\mathcal F_{PQ}(0)$. The relative
phase $-\pi/2$ gives the imaginary cross term
$ab\,\mathcal F_{PQ}(1)$. These are the two identities in
Eq.~\eqref{thm:families}. Since $ab=\sin(2\theta)/2$, their
normalization is the one in Eq.~\eqref{eq:angleACgenerators}.
The signed product axis satisfies $T^\dagger T=I$, so
$\lambda\mathbf S_T=2\lambda\mathcal D_T$.
\end{proof}

The family restriction is on the number of Pauli strings in a
single jump. It retains one- and two-string terms, and each string
can act on several qubits. The Hamiltonian family separately
supplies coherent rotations. The approximation order of products
using this restricted jump structure is established below for
single-qubit targets.

\section{Single-qubit compositions}\label{supp:sec:damping-example}

We measure how weak a target channel $\Phi$ is by
$\varepsilon=\|\Phi-\id\|_\diamond$, where the diamond norm is the
unhalved completely bounded trace norm. The main article uses the
normalized strength $p=\varepsilon/2$. All factors act on the
original qubit. We first allow signed strengths
and then require every factor to be a physical channel.

\begin{theorem}[Exact local representation with signed strengths]
\label{thm:weak-qubit-signed}
Let $\Phi$ be a single-qubit CPTP channel. Allow real $s$ in
the $\mathbf A$ and $\mathbf F$ families, with $\lambda=0$, and real
$\mathbf H$ rotation angles. For $\varepsilon<1$, $\Phi$ is the
exponential of a real sum of these generators and is a norm limit
of finite products of their exponentials. It also has an exact
finite-product representation, with the number of factors allowed to
depend on $\Phi$. In a neighborhood of
$\id$, $\Phi$ is an exact product of twelve such factors in a fixed
order, with parameters of order $\varepsilon$. Only the composite
is required to be CPTP.
\end{theorem}
\begin{proof}
For $\varepsilon<1$, the series for $\log\Phi$ in powers of
$\Phi-\id$ converges in the diamond norm and exponentiates to
$\Phi$ \cite[Sec.~II]{taxonomy2022}\cite[Sec.~11.3]{higham2008}.
Each positive power of $\Phi-\id$ preserves Hermiticity and
annihilates the trace. Thus $\log\Phi$ is TAHP, in the sense of
Theorem~\ref{lem:generatorbasis}.

Use the three $\mathbf H_P$ directions and one
$G_{\mathbf A}(0)=\mathbf S_P/2$ endpoint along each Pauli axis $P$.
For each of the three distinct pairs, also use
$G_{\mathbf A}(\pi/4)$ and $G_{\mathbf F}(\pi/4)$. The one-qubit
Pauli pairs anticommute, so these account for both cross directions
in Theorem~\ref{lem:generatorbasis}. At $\theta=\pi/4$ each complete
generator is $(\mathbf S_P+\mathbf S_Q)/4$ plus half its cross
direction. The twelve generators $G_j$ therefore form a real
basis by a triangular change of coordinates. Expanding $\log\Phi$
in this basis gives the sum. Theorem~\ref{thm:trotter-product}
gives its product limit, with convergence also in the diamond norm
by equivalence of norms in finite dimension.

For exact finite products, take the local product map
\[
F(b)=e^{b_{12}G_{12}}\circ\cdots\circ e^{b_1G_1}-\id .
\]
It is a smooth map from $\mathbb R^{12}$ into the real TAHP space,
and $DF(0)[h]=\sum_j h_j G_j$ is an isomorphism. The inverse function
theorem gives a smooth inverse on a neighborhood of zero, as in
second-kind exponential coordinates \cite[Lemma~3.2.9]{schmeding2024lie}.
The bounded derivative of that inverse on a sufficiently small
ball gives $\max_j|b_j|=O(\varepsilon)$.

For any $\varepsilon<1$, let $G=\log\Phi$ as above. For sufficiently
large integer $m$, the TAHP map $e^{G/m}-\id$ lies in the domain of
the local inverse. Thus $e^{G/m}$ has the ordered twelve-factor
representation. Repeating that product $m$ times gives
$(e^{G/m})^m=e^G=\Phi$.
\end{proof}

\begin{theorem}[Quadratic approximation with physical factors]
\label{thm:weak-qubit-physical}
For every single-qubit CPTP channel $\Phi$ and every $\delta>0$, there
exists a finite product $\Psi$ of physical $\mathbf H$ and $\mathbf A$
channels on that qubit such that
\begin{equation}
\|\Phi-\Psi\|_\diamond\le\tfrac12\varepsilon^2+\delta .
\label{eq:weak-qubit-quadratic}
\end{equation}
\end{theorem}

The proof combines the channel-difference generator of
Wolf and Cirac \cite[Lemma~1]{wolfcirac2008} with the single-qubit
jump reduction of Bacon et al.\ \cite[Theorem~3]{bacon2001}, also
used in the general composition framework of Sweke et al.\
\cite[Theorem~2]{sweke2015}.

\begin{proof}
Put $\Delta=\Phi-\id$ and choose Kraus operators $L_j$ for $\Phi$.
Trace preservation gives $\sum_jL_j^\dagger L_j=I$, so
$\Delta=\sum_j\mathcal D_{L_j}$. For $u\ge0$ the series
\[
e^{u\Delta}=e^{-u}\sum_{k\ge0}\frac{u^k}{k!}\Phi^k
\]
is a convex combination of CPTP maps and has diamond norm one.
Integrating the uniformly convergent series on $0\le u\le1$ gives
\[
e^\Delta-\Phi=\int_0^1(1-u)e^{u\Delta}\Delta^2\,du .
\]
Submultiplicativity yields
$\|e^\Delta-\Phi\|_\diamond\le\varepsilon^2/2$.

To approximate $e^\Delta$, write $L_j=c_jI+J_j$ with $J_j$ traceless.
Expansion gives $\mathcal D_{L_j}=\mathcal D_{J_j}-i[H_j,\cdot]$,
with the Hermitian operator $H_j=i(\overline{c_j}J_j-c_jJ_j^\dagger)/2$.
For a nonzero traceless jump $J=(x+iy)\cdot(X,Y,Z)$, an overall phase
rotates the real vectors $(x,y)$ and diagonalizes their real
$2\times2$ Gram matrix. After this rotation the vectors are
orthogonal and can be ordered with $\|x\|\ge\|y\|$. Single-qubit
unitary conjugation rotates the Pauli axes by an element of
$\mathrm{SO}(3)$, so the two unit directions are the images
of the $X$ and $Y$ axes under a common unitary $U$. If $y=0$,
choose a perpendicular direction for the unused axis. Thus
\[
J=a e^{i\alpha}U(\cos\chi\,X+i\sin\chi\,Y)U^\dagger ,
\qquad a^2=\|x\|^2+\|y\|^2,\quad 0\le\chi\le\pi/4 .
\]
The overall phase cancels from the dissipator. Proposition~
\ref{prop:complete-jump-family-restriction} gives
\begin{equation}
e^{\mathcal D_J}=
\C_U\circ\N_{\mathbf A}(a^2,-\chi,0)\circ\C_{U^\dagger}.
\label{eq:qubit-jump-products}
\end{equation}
Zero jumps contribute the identity. Both frame rotations and the
exponential of the summed Hamiltonian are finite $\mathbf H$
products by the single-qubit Euler factorization into rotations
about $Z,Y,Z$. Apply Theorem~\ref{thm:trotter-product} to the
Hamiltonian term and the traceless dissipators. In each step
dividing a dissipator by the number of steps divides $a^2$ in
Eq.~\eqref{eq:qubit-jump-products} by the same number, so its
strength remains nonnegative. The resulting physical products
converge to $e^\Delta$ in diamond norm. Choosing a product within
$\delta$ of $e^\Delta$ proves \eqref{eq:weak-qubit-quadratic}.
\end{proof}

For a general weak target, Theorem~\ref{thm:weak-qubit-signed} expands
$\log\Phi$ in the family generators, and
Theorem~\ref{thm:weak-qubit-physical} uses $\Phi-\id$ to obtain a
quadratic error bound with physical factors. For decay to
$\lvert0\rangle$ with a complex coherence multiplier, the same jump
reduction in Eq.~\eqref{eq:qubit-jump-products} gives an exact physical
product with commuting generators. The population and coherence data
determine all three strengths directly.

\begin{proposition}[Exact fit for decay, dephasing and phase rotation]
Let a target $\Phi$ act in the ordered $Z$ eigenbasis $\lvert0\rangle,
\lvert1\rangle$ as
\begin{equation}
\Phi(\rho)=
\begin{pmatrix}
\rho_{++}+(1-z)\rho_{--} & u\rho_{+-}\\
\overline u\rho_{-+} & z\rho_{--}
\end{pmatrix},
\label{eq:qubit-composed-blocks}
\end{equation}
where $0<z\le1$ and $0<|u|\le\sqrt z$. Here $z$ is the population
retained by $\lvert1\rangle$ from that input and $u$ is the multiplier
of the $\lvert0\rangle\!\langle1\rvert$ coherence. With axes $P=X$,
$Q=Y$ and $T=Z$, the target equals
\begin{equation}
\Phi=\N_{\mathbf H}(\theta)\circ\N_{\mathbf S}(\lambda)\circ
\N_{\mathbf A}(s/2,-\pi/4,0),
\label{supp:eq:qubit-composed-noise}
\end{equation}
where
\[
s=-\log z,\qquad
\lambda=\tfrac14\log\frac{\sqrt z}{|u|},\qquad
\theta=-\tfrac12\arg u\pmod{\pi}.
\]
\end{proposition}
\begin{proof}
The amplitude-damping jump $\sqrt{s}\,\lvert0\rangle\!\langle1\rvert
=\sqrt{s}(X+iY)/2$ is the case $a^2=s/2$, $\chi=\pi/4$ and $U=I$
in Eq.~\eqref{eq:qubit-jump-products}. The three generators preserve
the diagonal subspace and each of the two one-dimensional coherence
subspaces. On the diagonal subspace only the damping generator acts.
On each coherence subspace all three generators act by scalars.
Thus the generators commute and the product is the exponential of
their sum.

Equation~\eqref{eq:ampchannel} gives population survival $e^{-s}$ and
coherence attenuation $e^{-s/2}$. The $Z$ dephasing supplies the
remaining attenuation $e^{-4\lambda}$ and the rotation supplies the
phase $e^{-2i\theta}$ on the upper off-diagonal entry by
Eq.~\eqref{eq:singleHblocks}. Equating the resulting population and
coherence factors with $z$ and $u$ gives the stated strengths, with
$s,\lambda\ge0$.
\end{proof}

\section{Non-unital channels}\label{supp:sec:bias}

A change in the joint maximally mixed state can appear in a parity measurement even when both qubits remain maximally mixed individually. The action of each generator on the identity determines which classes can produce this effect.

A channel $\Gamma$ on $n$ qubits is unital if $\Gamma(I)=I$, equivalently if $\Gamma(I/2^n)=I/2^n$.

\begin{proposition}[Identity action and non-unital channels]\label{prop:identity-action}
Random Pauli channels are unital. Of the six classes, only $\mathbf{A}$ acts nontrivially on $I$. For a complete $\mathbf{A}$ generator with parameters $s,\theta,\lambda$ as in Table~\ref{tab:classes},
\[
\N_{\mathbf{A}}(s,\theta,\lambda)(I)=I-(1-e^{-2s})\sin2\theta\,T.
\]
Thus only the $\mathbf{A}$ family can give a non-unital channel in the table. In a sum of generators, distinct $\mathbf{A}$ terms can cancel on $I$.
\end{proposition}
\begin{proof}
Each Pauli conjugation fixes $I$, hence so does its probability mixture. For the basis of Theorem~\ref{lem:generatorbasis}, direct substitution gives
\begin{align}
\mathcal F_{PQ}(\alpha)(I)&=(i^\alpha-i^{-\alpha})[P,Q],\notag\\
\mathbf{H}_P(I)=\mathbf{S}_P(I)=\mathbf{C}_{P,Q}(I)=\mathbf{E}_{P,Q}(I)=\mathbf{F}_{P,Q}(I)&=0,\qquad
\mathbf{A}_{P,Q}(I)=4iPQ=-4T .
\label{eq:classidentity}
\end{align}
For any generator $G$ with $G(I)=0$, all positive powers of $G$ annihilate $I$. Hence
\begin{equation}
G(I)=0\quad\Longrightarrow\quad e^G(I)=I .
\label{eq:unitalgenerator}
\end{equation}
This proves unitality for $\mathbf{H}$, $\mathbf{S}$, $\mathbf{C}$, $\mathbf{E}$ and $\mathbf{F}$. In the $\mathbf{A}$ block map, the input $I$ has diagonal blocks $\Pi_\sigma$. The output coefficient is $p_\sigma+q_{-\sigma}=1-\sigma(q_+-q_-)$, which gives the asserted formula. The pairs $(X_1,Y_1)$ and $(X_1Z_2,Y_1Z_2)$ both have $T=Z_1$. Their complete $\mathbf{A}$ generators at equal strengths and opposite angles $\pm\pi/4$ act on $I$ with equal magnitudes and opposite signs.
\end{proof}

\begin{proposition}[Joint parity with maximally mixed marginals]\label{prop:pair-parity}
Choose $P=X_1$, $Q=Y_1Z_2$ and $\lambda=0$. For $s\ge0$ and $\theta\in\mathbb R$,
\begin{equation}
\N_{\mathbf{A}}(s,\theta,0)(I/4)
=\tfrac14\bigl[I-(1-e^{-2s})\sin 2\theta\, Z_1Z_2\bigr].
\label{supp:eq:two-qubit-parity-bias}
\end{equation}
A $Z_1Z_2$ parity measurement has unequal outcome probabilities exactly when $s>0$ and $\sin 2\theta\ne0$. Both single-qubit marginals are $I/2$.
\end{proposition}
\begin{proof}
The signed axis for this pair is $T=Z_1Z_2$. Linearity and Proposition~\ref{prop:identity-action} give the displayed output. The two eigenspaces have rank two by Proposition~\ref{prop:pauli-sector-geometry}, so the even and odd parity probabilities are $(1\mp(1-e^{-2s})\sin2\theta)/2$, respectively. The term $Z_1Z_2$ has zero partial trace on each qubit because $\tr Z=0$.
\end{proof}

\section{Leakage and correlated disturbances}\label{supp:sec:leakage}

We apply the channel families to a four-level system with computational and leakage subspaces. After choosing coordinates for the internal state and the occupied subspace, we construct channels for leakage and return, dephasing during leakage, and rotations of a neighboring qubit at a transition. We then model readout and reset, identify the conditions for a known-location erasure, and show how persistent leakage can correlate errors at different circuit locations.

\subsection{Subspace representation}

Following leakage across circuit operations requires keeping track of the occupied subspace and the state within it. The four-level model assigns these roles to two coordinates of the same system.

\begin{definition}[Effective leakage model]\label{def:leakagemodel}
Let $\mathcal H_{\rm C}$ and $\mathcal H_{\rm L}$ be orthogonal subspaces of dimension two, with chosen orthonormal bases $|j\rangle_{\rm C}$ and $|j\rangle_{\rm L}$, $j=0,1$. The internal coordinate $d$ and leakage register $v$ are defined by
\begin{align}
\mathcal H_{\rm C}\oplus\mathcal H_{\rm L}&\simeq\mathbb C_d^2\otimes\mathbb C_v^2,\notag\\
|j\rangle_{\rm C}&\mapsto |j\rangle_d|0\rangle_v,\qquad
|j\rangle_{\rm L}\mapsto |j\rangle_d|1\rangle_v.
\label{eq:leakage-sector-isomorphism}
\end{align}
Leakage is $v:0\to1$ and return is $v:1\to0$. Write $\Pi_b^{(v)}$ for the projector onto $v=b$, with identity on all spectator systems.
\end{definition}

\begin{proposition}[Leakage coordinates and the block-diagonal restriction]\label{prop:leakage-sector-representation}
The identification in Eq.~\eqref{eq:leakage-sector-isomorphism} is unitary. It transports states and channel actions without changing their trace, positivity or compositions, including on extensions by an identity on a reference. A state block diagonal in $v$ is specified by the occupation probabilities and conditional states. When the channels preserve this block structure, a classical occupation label and the corresponding conditional state suffice to represent their outputs.
\end{proposition}
\begin{proof}
The identification maps one orthonormal basis onto another, so it is a unitary change of coordinates on the same four-dimensional system. Conjugation by this unitary, extended by the identity on all other systems, preserves traces and positivity and intertwines compositions. In this representation
\[
X=\sum_{b,c=0}^1 X_{bc}\otimes |b\rangle\langle c|_v,
\qquad X_{bc}=\langle b|X|c\rangle_v .
\]
For a block-diagonal density operator, the probabilities are $p_b=\tr X_{bb}$ and the conditional states are $X_{bb}/p_b$ when $p_b>0$. A zero-probability branch has $X_{bb}=0$ by positivity. The joint state is recovered by summing the weighted conditional states tensored with their orthogonal occupation flags. If a channel preserves this form, the same representation applies at its output. A computational input has support at $b=0$.
\end{proof}

\subsection{Subspace transfer and conditional disturbances}\label{sec:flag-noise-coverage}

The occupation coordinate lets us describe three effects in the same representation. It can change during leakage and return, control dephasing while the system is leaked, and switch in the same event that rotates a neighboring qubit.

\begin{proposition}[Directed leakage and return]\label{prop:flagcontrols}
For $P=I_d\otimes X_v$, $Q=I_d\otimes Y_v$, $T=I_d\otimes Z_v$, the channels $\N_{\mathbf{A}}(\ell/2,\pi/4,0)$ and $\N_{\mathbf{A}}(r/2,-\pi/4,0)$ with $\ell,r\ge0$ produce one-way leakage and one-way return, respectively. Their transition probabilities from the respective source subspaces are $1-e^{-\ell}$ and $1-e^{-r}$. Both carry the current internal state unchanged between the paired subspaces.
\end{proposition}
\begin{proof}
The jumps $L_\ell=I_d\otimes |1\rangle\langle0|_v$ and $L_r=I_d\otimes |0\rangle\langle1|_v$ satisfy
\begin{equation}
\mathcal D_{L_\ell}=\tfrac12G_{\mathbf{A}}(\pi/4),\qquad
\mathcal D_{L_r}=\tfrac12G_{\mathbf{A}}(-\pi/4).
\label{eq:flag-A-endpoints}
\end{equation}
Indeed, the jumps of $G_{\mathbf{A}}(\pm\pi/4)$ are $(P\mp iQ)/\sqrt2=\sqrt2 L_{\ell,r}$. Substituting $\ell/2$ or $r/2$ into Eq.~\eqref{eq:angleACweights} gives the stated probabilities and zero transfer in the opposite direction for each endpoint. The Kraus operators act only on $v$, so each transferred block is a scalar multiple of the corresponding source block, including its internal and reference state. A subspace is absorbing under a collection of channels when every channel in that collection sends states supported there to states supported there. One-way leakage has this property on $\mathcal H_{\rm L}$, as does any composition with other channels that preserve $\mathcal H_{\rm L}$. A return transfers the internal block in its input.
\end{proof}

\begin{proposition}[Noise conditional on leaked occupation]\label{prop:flag-noise-components}
\label{thm:leakage}
For $s\ge0$ and a nonidentity Pauli $R$ on the internal or a neighboring qubit, set $P=R\otimes I_v$, $Q=R\otimes Z_v$, with $T=I\otimes Z_v$. Then
\begin{equation}
\mathcal D_{R\otimes\proj1_v}=\tfrac12G_{\mathbf{C}}(-\pi/4),\qquad
\N_{\mathbf{C}}(s/2,-\pi/4,0)=e^{s\mathcal D_{R\otimes\proj1_v}}.
\label{eq:flag-C-residence}
\end{equation}
The $00$ block is unchanged, the $11$ block becomes
\begin{equation}
(\N_{\mathbf{C}}(s/2,-\pi/4,0)(X))_{11}
=\tfrac{1+e^{-2s}}2 X_{11}+\tfrac{1-e^{-2s}}2 RX_{11}R ,
\label{eq:flag-C-residence-map}
\end{equation}
and each cross block is multiplied by $e^{-s/2}$. The channel preserves both subspace populations and can correlate the affected qubit with the occupation $v$. Compose the $R=X_d$ and $R=Z_d$ choices with family arguments $s_X/2$ and $s_Z/2$, respectively, and with the same angle and added dephasing as above. As $s_X,s_Z\to\infty$, the composition fixes the $00$ block, removes the off-diagonal blocks, and has
\begin{equation}
X_{11}\longmapsto \frac{I_d}{2}\otimes\tr_d X_{11}.
\label{eq:erasure}
\end{equation}
\end{proposition}
\begin{proof}
The $\mathbf{C}$ jump at $-\pi/4$ is $(P-Q)/\sqrt2=\sqrt2 R\otimes\proj1_v$. Equation~\eqref{eq:canonicalKblocks} at family strength $s/2$ gives the stated block factors. Equivalently, the conditional jump has squared modulus $\Pi_1^{(v)}$, so its dissipator acts by $\C_R-\id$ on $X_{11}$ and by $-1/2$ on cross blocks. One axis preserves its eigenstate populations even at infinite strength. The channel preserves the population of each subspace and can correlate that occupation with the affected qubit. For example, take an affected qubit in $\proj0$ and an independent leakage register in a diagonal state with both values populated. With $R=X$ and $s>0$, the probability of finding the affected qubit in $|1\rangle$ is zero given $v=0$ and $(1-e^{-2s})/2$ given $v=1$. The joint state therefore develops a correlation while staying block diagonal.

For the composed $X_d$ and $Z_d$ choices the $11$ block tends to the product of their equal-weight dephasing maps. The conjugations commute, and their product is conjugation by $Y_d$ because the scalar phase in $X_dZ_d$ cancels. The limit on that block is the uniform average of the four Pauli conjugations on $d$. Expanding in the data Pauli basis with operator coefficients on the other systems shows each traceless Pauli term cancels and the identity term is $I_d/2\otimes\tr_d X_{11}$. This proves Eq.~\eqref{eq:erasure} also for an entangled reference. On the $11$ block the $X_d$ component is attenuated by $e^{-2s_Z}$ and the $Z_d$ component by $e^{-2s_X}$, so complete loss for arbitrary inputs requires both strengths to diverge. The cross-block scalar is $e^{-(s_X+s_Z)/2}$ and vanishes. The block coefficients therefore converge in norm. The limiting complete channel is CPTP, and its leaked block has lost both the internal information and its conditional correlations with the other systems.
\end{proof}

\begin{proposition}[Neighbor rotation coupled to a subspace change]
For a neighboring qubit $a$ distinct from $d$, and a nonidentity Pauli $R_a$, choose $P=I_a\otimes X_v$ and $Q=R_a\otimes X_v$, with identities on all other systems. Then $T=R_a\otimes I_v$ and
\begin{equation}
U_\theta=e^{-i\theta R_a}\otimes X_v .
\label{eq:neighbor-E-event}
\end{equation}
For $s\ge0$, $\theta\in\mathbb R$ and input occupation $b\in\{0,1\}$, the Poisson count $m$ labels a branch with output occupation $b$ for even $m$ and $1-b$ for odd $m$. The switching probability is $(1-e^{-2s})/2$ for either input. The neighbor is conjugated by $e^{-im\theta R_a}$ in branch $m$. An ideal occupation report for a known input occupation identifies the parity of $m$. The conditioned neighbor state can differ from its input state.
\end{proposition}
\begin{proof}
By Proposition~\ref{lem:rotatingpulses}, $U_\theta^m=e^{-im\theta R_a}\otimes X_v^m$, with weight $p_m=e^{-s}s^m/m!$. Summing the odd weights gives $e^{-s}\sinh s=(1-e^{-2s})/2$. Thus an ideal occupation report and a known input occupation identify the parity of $m$. For $s>0$, both counts $0$ and $2$ have positive weight and the same parity. For $R_a=Z_a$ and $\theta=\pi/4$, their rotations send $|+\rangle_a=(|0\rangle_a+|1\rangle_a)/\sqrt2$ to opposite $X_a$ eigenstates up to global phases.

The rotation can be visible in the neighbor marginal selected by a leakage report. With the same $R_a$ and $\theta$, take the input $|+\rangle_a$ and $v=0$. The unnormalized expectation of $Y_a$ in the reported leaked block is
\begin{equation}
\sum_{m\ {\rm odd}}p_m\sin(m\pi/2)=e^{-s}\sin s,
\label{eq:neighbor-E-witness}
\end{equation}
by the sine series. This is nonzero for $0<s<\pi$, whereas the input expectation is zero. If the internal coordinate $d$ is erased by Eq.~\eqref{eq:erasure}, partial trace over $d$ gives the same neighbor marginal in that block.
\end{proof}

\subsection{Leakage readout, reset and shared occupation}\label{supp:sec:flag-records}

In the erasure limit, an ideal occupation report identifies the branch that has lost its encoded information. The same register can also correlate errors at different circuit locations when its occupation persists.

In an ideal computational-basis readout of $v$, report $b\in\{0,1\}$ selects the branch
\begin{equation}
\mathcal M_b(X)=\Pi_b^{(v)}X\Pi_b^{(v)}.
\label{eq:flag-reported-instrument}
\end{equation}
Each branch has one Kraus operator and occurs with probability $\tr\mathcal M_b(X)$ for a normalized input. Discarding the report sums the two branches, which removes the cross blocks and preserves the diagonal blocks and their total trace.

A reset triggered by report $1$ gives the branches
\begin{equation}
\mathcal J_0=\mathcal M_0,\qquad
\mathcal J_1=\mathcal R_{00}\circ\mathcal M_1,\qquad
\mathcal R_{00}(X)=\proj{00}_{dv}\otimes\tr_{dv}X.
\label{eq:reported-leakage-reset}
\end{equation}
The reset discards $(d,v)$ and prepares $|00\rangle_{dv}$, preserving the probability of the branch and the reduced state on all other systems within that branch.

\begin{proposition}[$\mathbf{A}$/$\mathbf{C}$ realization of known-location erasure, using the flag convention of Ref.~{\cite{wu2022}}]
\label{prop:reported-erasure}
Take a normalized computational input $\rho_{d\mathcal R}\otimes\proj0_v$ with any reference $\mathcal R$. Compose one-way $\mathbf{A}$ leakage with parameter $\ell\ge0$, the complete conditional loss on $d$ in Eq.~\eqref{eq:erasure}, and ideal readout. The sum of the two report branches is
\begin{equation}
(1-p)\rho_{d\mathcal R}\otimes\proj0_v
+p\,\frac{I_d}{2}\otimes\tr_d(\rho_{d\mathcal R})\otimes\proj1_v,
\qquad p=1-e^{-\ell}.
\label{eq:flag-erasure-conversion}
\end{equation}
The report distinguishes the two branches. The erasure probability $p$ is independent of $\rho_{d\mathcal R}$, and the result includes complete erasure in the limit $\ell\to\infty$.
\end{proposition}
\begin{proof}
Proposition~\ref{prop:flagcontrols} assigns probabilities $1-p,p$ to the two flags and preserves $\rho_{d\mathcal R}$ in both conditional branches of nonzero weight. Equation~\eqref{eq:erasure} fixes the $v=0$ block and replaces the $d$ part of the $v=1$ block by $I_d/2$, leaving its reference marginal. The ideal readout in Eq.~\eqref{eq:flag-reported-instrument} labels the resulting branches. At $\ell=0$ the leaked branch has zero weight. Single-axis dephasing can retain information in the leaked block.
\end{proof}

\begin{proposition}[Correlations from shared leakage occupation]
\label{prop:flagmemory}
The reduced data output of a channel controlled by $v$ need not be determined by the data marginal alone. In particular, let two data qubits in $|00\rangle$ share a diagonal leakage register with probability $w\in[0,1]$ of $v=1$. Use conditional $\mathbf{C}$ dephasing along $X$ on each qubit with strengths $s_1,s_2\ge0$ in Eq.~\eqref{eq:flag-C-residence-map}, and measure both in the $Z$ basis. For the resulting binary outcomes $B_1,B_2$, let $a_j=(1-e^{-2s_j})/2$. Then
\begin{equation}
\operatorname{Cov}(B_1,B_2)=w(1-w)a_1a_2 .
\label{eq:shared-flag-correlation}
\end{equation}
Discarding an occupation report leaves the joint state in this example unchanged.
\end{proposition}
\begin{proof}
A data qubit in $|0\rangle$ gives different outputs for $v=0$ and $v=1$ under conditional $X$ dephasing of positive strength. For $v=0$ the channel fixes the input, while for $v=1$ the probability of output $|1\rangle$ is positive by Eq.~\eqref{eq:flag-C-residence-map}. These joint inputs have the same data marginal.

In the two-qubit example, both measurement outcomes are zero conditioned on $v=0$. Conditioned on $v=1$, the two maps act independently on the product input, giving independent outcomes with probabilities $a_1,a_2$ of $1$. Therefore $\mathbb E[B_j]=wa_j$ and $\mathbb E[B_1B_2]=wa_1a_2$, which give the stated covariance. The joint state is block diagonal in $v$, so summing Eq.~\eqref{eq:flag-reported-instrument} over $b$ preserves it.
\end{proof}


\begin{thebibliography}{24}%
\makeatletter
\providecommand \@ifxundefined [1]{%
 \@ifx{#1\undefined}
}%
\providecommand \@ifnum [1]{%
 \ifnum #1\expandafter \@firstoftwo
 \else \expandafter \@secondoftwo
 \fi
}%
\providecommand \@ifx [1]{%
 \ifx #1\expandafter \@firstoftwo
 \else \expandafter \@secondoftwo
 \fi
}%
\providecommand \natexlab [1]{#1}%
\providecommand \enquote  [1]{``#1''}%
\providecommand \bibnamefont  [1]{#1}%
\providecommand \bibfnamefont [1]{#1}%
\providecommand \citenamefont [1]{#1}%
\providecommand \href@noop [0]{\@secondoftwo}%
\providecommand \href [0]{\begingroup \@sanitize@url \@href}%
\providecommand \@href[1]{\@@startlink{#1}\@@href}%
\providecommand \@@href[1]{\endgroup#1\@@endlink}%
\providecommand \@sanitize@url [0]{\catcode `\\12\catcode `\$12\catcode
  `\&12\catcode `\#12\catcode `\^12\catcode `\_12\catcode `\%12\relax}%
\providecommand \@@startlink[1]{}%
\providecommand \@@endlink[0]{}%
\providecommand \url  [0]{\begingroup\@sanitize@url \@url }%
\providecommand \@url [1]{\endgroup\@href {#1}{\urlprefix }}%
\providecommand \urlprefix  [0]{URL }%
\providecommand \Eprint [0]{\href }%
\providecommand \doibase [0]{https://doi.org/}%
\providecommand \selectlanguage [0]{\@gobble}%
\providecommand \bibinfo  [0]{\@secondoftwo}%
\providecommand \bibfield  [0]{\@secondoftwo}%
\providecommand \translation [1]{[#1]}%
\providecommand \BibitemOpen [0]{}%
\providecommand \bibitemStop [0]{}%
\providecommand \bibitemNoStop [0]{.\EOS\space}%
\providecommand \EOS [0]{\spacefactor3000\relax}%
\providecommand \BibitemShut  [1]{\csname bibitem#1\endcsname}%
\let\auto@bib@innerbib\@empty
\bibitem [{\citenamefont {Bravyi}\ \emph {et~al.}(2018)\citenamefont {Bravyi},
  \citenamefont {Englbrecht}, \citenamefont {König},\ and\ \citenamefont
  {Peard}}]{bravyi2018}%
  \BibitemOpen
  \bibfield  {author} {\bibinfo {author} {\bibfnamefont {S.}~\bibnamefont
  {Bravyi}}, \bibinfo {author} {\bibfnamefont {M.}~\bibnamefont {Englbrecht}},
  \bibinfo {author} {\bibfnamefont {R.}~\bibnamefont {König}},\ and\ \bibinfo
  {author} {\bibfnamefont {N.}~\bibnamefont {Peard}},\ }\bibfield  {title}
  {\bibinfo {title} {{Correcting coherent errors with surface codes}},\ }\href
  {https://doi.org/10.1038/s41534-018-0106-y} {\bibfield  {journal} {\bibinfo
  {journal} {npj Quantum Information}\ }\textbf {\bibinfo {volume} {4}},\
  \bibinfo {pages} {55} (\bibinfo {year} {2018})},\ \Eprint
  {https://arxiv.org/abs/1710.02270} {arXiv:1710.02270 [quant-ph]} \BibitemShut
  {NoStop}%
\bibitem [{\citenamefont {Bonilla~Ataides}\ \emph {et~al.}(2021)\citenamefont
  {Bonilla~Ataides}, \citenamefont {Tuckett}, \citenamefont {Bartlett},
  \citenamefont {Flammia},\ and\ \citenamefont {Brown}}]{bonilla2021}%
  \BibitemOpen
  \bibfield  {author} {\bibinfo {author} {\bibfnamefont {J.~P.}\ \bibnamefont
  {Bonilla~Ataides}}, \bibinfo {author} {\bibfnamefont {D.~K.}\ \bibnamefont
  {Tuckett}}, \bibinfo {author} {\bibfnamefont {S.~D.}\ \bibnamefont
  {Bartlett}}, \bibinfo {author} {\bibfnamefont {S.~T.}\ \bibnamefont
  {Flammia}},\ and\ \bibinfo {author} {\bibfnamefont {B.~J.}\ \bibnamefont
  {Brown}},\ }\bibfield  {title} {\bibinfo {title} {The {XZZX} surface code},\
  }\href {https://doi.org/10.1038/s41467-021-22274-1} {\bibfield  {journal}
  {\bibinfo  {journal} {Nature Communications}\ }\textbf {\bibinfo {volume}
  {12}},\ \bibinfo {pages} {2172} (\bibinfo {year} {2021})},\ \Eprint
  {https://arxiv.org/abs/2009.07851} {arXiv:2009.07851 [quant-ph]} \BibitemShut
  {NoStop}%
\bibitem [{\citenamefont {Dennis}\ \emph {et~al.}(2002)\citenamefont {Dennis},
  \citenamefont {Kitaev}, \citenamefont {Landahl},\ and\ \citenamefont
  {Preskill}}]{dennis2002}%
  \BibitemOpen
  \bibfield  {author} {\bibinfo {author} {\bibfnamefont {E.}~\bibnamefont
  {Dennis}}, \bibinfo {author} {\bibfnamefont {A.}~\bibnamefont {Kitaev}},
  \bibinfo {author} {\bibfnamefont {A.}~\bibnamefont {Landahl}},\ and\ \bibinfo
  {author} {\bibfnamefont {J.}~\bibnamefont {Preskill}},\ }\bibfield  {title}
  {\bibinfo {title} {{Topological quantum memory}},\ }\href
  {https://doi.org/10.1063/1.1499754} {\bibfield  {journal} {\bibinfo
  {journal} {Journal of Mathematical Physics}\ }\textbf {\bibinfo {volume}
  {43}},\ \bibinfo {pages} {4452} (\bibinfo {year} {2002})},\ \Eprint
  {https://arxiv.org/abs/quant-ph/0110143} {arXiv:quant-ph/0110143}
  \BibitemShut {NoStop}%
\bibitem [{\citenamefont {Bausch}\ \emph {et~al.}(2024)\citenamefont {Bausch},
  \citenamefont {Senior}, \citenamefont {Heras}, \citenamefont {Edlich},
  \citenamefont {Davies}, \citenamefont {Newman}, \citenamefont {Jones},
  \citenamefont {Satzinger}, \citenamefont {Niu}, \citenamefont {Blackwell},
  \citenamefont {Holland}, \citenamefont {Kafri}, \citenamefont {Atalaya},
  \citenamefont {Gidney}, \citenamefont {Hassabis}, \citenamefont {Boixo},
  \citenamefont {Neven},\ and\ \citenamefont {Kohli}}]{bausch2024}%
  \BibitemOpen
  \bibfield  {author} {\bibinfo {author} {\bibfnamefont {J.}~\bibnamefont
  {Bausch}}, \bibinfo {author} {\bibfnamefont {A.~W.}\ \bibnamefont {Senior}},
  \bibinfo {author} {\bibfnamefont {F.~J.~H.}\ \bibnamefont {Heras}}, \bibinfo
  {author} {\bibfnamefont {T.}~\bibnamefont {Edlich}}, \bibinfo {author}
  {\bibfnamefont {A.}~\bibnamefont {Davies}}, \bibinfo {author} {\bibfnamefont
  {M.}~\bibnamefont {Newman}}, \bibinfo {author} {\bibfnamefont
  {C.}~\bibnamefont {Jones}}, \bibinfo {author} {\bibfnamefont
  {K.}~\bibnamefont {Satzinger}}, \bibinfo {author} {\bibfnamefont {M.~Y.}\
  \bibnamefont {Niu}}, \bibinfo {author} {\bibfnamefont {S.}~\bibnamefont
  {Blackwell}}, \bibinfo {author} {\bibfnamefont {G.}~\bibnamefont {Holland}},
  \bibinfo {author} {\bibfnamefont {D.}~\bibnamefont {Kafri}}, \bibinfo
  {author} {\bibfnamefont {J.}~\bibnamefont {Atalaya}}, \bibinfo {author}
  {\bibfnamefont {C.}~\bibnamefont {Gidney}}, \bibinfo {author} {\bibfnamefont
  {D.}~\bibnamefont {Hassabis}}, \bibinfo {author} {\bibfnamefont
  {S.}~\bibnamefont {Boixo}}, \bibinfo {author} {\bibfnamefont
  {H.}~\bibnamefont {Neven}},\ and\ \bibinfo {author} {\bibfnamefont
  {P.}~\bibnamefont {Kohli}},\ }\bibfield  {title} {\bibinfo {title} {{Learning
  high-accuracy error decoding for quantum processors}},\ }\href
  {https://doi.org/10.1038/s41586-024-08148-8} {\bibfield  {journal} {\bibinfo
  {journal} {Nature}\ }\textbf {\bibinfo {volume} {635}},\ \bibinfo {pages}
  {834} (\bibinfo {year} {2024})}\BibitemShut {NoStop}%
\bibitem [{\citenamefont {Varbanov}\ \emph {et~al.}(2020)\citenamefont
  {Varbanov}, \citenamefont {Battistel}, \citenamefont {Tarasinski},
  \citenamefont {Ostroukh}, \citenamefont {O'Brien}, \citenamefont {DiCarlo},\
  and\ \citenamefont {Terhal}}]{varbanov2020}%
  \BibitemOpen
  \bibfield  {author} {\bibinfo {author} {\bibfnamefont {B.~M.}\ \bibnamefont
  {Varbanov}}, \bibinfo {author} {\bibfnamefont {F.}~\bibnamefont {Battistel}},
  \bibinfo {author} {\bibfnamefont {B.~M.}\ \bibnamefont {Tarasinski}},
  \bibinfo {author} {\bibfnamefont {V.~P.}\ \bibnamefont {Ostroukh}}, \bibinfo
  {author} {\bibfnamefont {T.~E.}\ \bibnamefont {O'Brien}}, \bibinfo {author}
  {\bibfnamefont {L.}~\bibnamefont {DiCarlo}},\ and\ \bibinfo {author}
  {\bibfnamefont {B.~M.}\ \bibnamefont {Terhal}},\ }\bibfield  {title}
  {\bibinfo {title} {{Leakage detection for a transmon-based surface code}},\
  }\href {https://doi.org/10.1038/s41534-020-00330-w} {\bibfield  {journal}
  {\bibinfo  {journal} {npj Quantum Information}\ }\textbf {\bibinfo {volume}
  {6}},\ \bibinfo {pages} {102} (\bibinfo {year} {2020})}\BibitemShut {NoStop}%
\bibitem [{\citenamefont {Blume-Kohout}\ \emph {et~al.}(2022)\citenamefont
  {Blume-Kohout}, \citenamefont {da~Silva}, \citenamefont {Nielsen},
  \citenamefont {Proctor}, \citenamefont {Rudinger}, \citenamefont {Sarovar},\
  and\ \citenamefont {Young}}]{taxonomy2022}%
  \BibitemOpen
  \bibfield  {author} {\bibinfo {author} {\bibfnamefont {R.}~\bibnamefont
  {Blume-Kohout}}, \bibinfo {author} {\bibfnamefont {M.~P.}\ \bibnamefont
  {da~Silva}}, \bibinfo {author} {\bibfnamefont {E.}~\bibnamefont {Nielsen}},
  \bibinfo {author} {\bibfnamefont {T.}~\bibnamefont {Proctor}}, \bibinfo
  {author} {\bibfnamefont {K.}~\bibnamefont {Rudinger}}, \bibinfo {author}
  {\bibfnamefont {M.}~\bibnamefont {Sarovar}},\ and\ \bibinfo {author}
  {\bibfnamefont {K.}~\bibnamefont {Young}},\ }\bibfield  {title} {\bibinfo
  {title} {{A Taxonomy of Small Markovian Errors}},\ }\href
  {https://doi.org/10.1103/prxquantum.3.020335} {\bibfield  {journal} {\bibinfo
   {journal} {PRX Quantum}\ }\textbf {\bibinfo {volume} {3}},\ \bibinfo {pages}
  {020335} (\bibinfo {year} {2022})},\ \Eprint
  {https://arxiv.org/abs/2103.01928} {arXiv:2103.01928 [quant-ph]} \BibitemShut
  {NoStop}%
\bibitem [{\citenamefont {Miller}\ \emph {et~al.}(2025)\citenamefont {Miller},
  \citenamefont {Ostrove}, \citenamefont {Hines}, \citenamefont {Blume-Kohout},
  \citenamefont {Young},\ and\ \citenamefont {Proctor}}]{miller2025}%
  \BibitemOpen
  \bibfield  {author} {\bibinfo {author} {\bibfnamefont {A.}~\bibnamefont
  {Miller}}, \bibinfo {author} {\bibfnamefont {C.}~\bibnamefont {Ostrove}},
  \bibinfo {author} {\bibfnamefont {J.}~\bibnamefont {Hines}}, \bibinfo
  {author} {\bibfnamefont {R.}~\bibnamefont {Blume-Kohout}}, \bibinfo {author}
  {\bibfnamefont {K.}~\bibnamefont {Young}},\ and\ \bibinfo {author}
  {\bibfnamefont {T.}~\bibnamefont {Proctor}},\ }\href@noop {} {\bibinfo
  {title} {{Efficient simulation of Clifford circuits with small Markovian
  errors}}} (\bibinfo {year} {2025}),\ \bibinfo {note} {preprint},\ \Eprint
  {https://arxiv.org/abs/2504.15128v1} {arXiv:2504.15128v1 [quant-ph]}
  \BibitemShut {NoStop}%
\bibitem [{\citenamefont {Trotter}(1959)}]{trotter1959}%
  \BibitemOpen
  \bibfield  {author} {\bibinfo {author} {\bibfnamefont {H.~F.}\ \bibnamefont
  {Trotter}},\ }\bibfield  {title} {\bibinfo {title} {On the product of
  semi-groups of operators},\ }\href
  {https://doi.org/10.1090/S0002-9939-1959-0108732-6} {\bibfield  {journal}
  {\bibinfo  {journal} {Proceedings of the American Mathematical Society}\
  }\textbf {\bibinfo {volume} {10}},\ \bibinfo {pages} {545} (\bibinfo {year}
  {1959})}\BibitemShut {NoStop}%
\bibitem [{\citenamefont {Wood}\ and\ \citenamefont
  {Gambetta}(2018)}]{wood2018}%
  \BibitemOpen
  \bibfield  {author} {\bibinfo {author} {\bibfnamefont {C.~J.}\ \bibnamefont
  {Wood}}\ and\ \bibinfo {author} {\bibfnamefont {J.~M.}\ \bibnamefont
  {Gambetta}},\ }\bibfield  {title} {\bibinfo {title} {{Quantification and
  Characterization of Leakage Errors}},\ }\href
  {https://doi.org/10.1103/PhysRevA.97.032306} {\bibfield  {journal} {\bibinfo
  {journal} {Physical Review A}\ }\textbf {\bibinfo {volume} {97}},\ \bibinfo
  {pages} {032306} (\bibinfo {year} {2018})},\ \Eprint
  {https://arxiv.org/abs/1704.03081} {arXiv:1704.03081 [quant-ph]} \BibitemShut
  {NoStop}%
\bibitem [{\citenamefont {Suchara}\ \emph {et~al.}(2015)\citenamefont
  {Suchara}, \citenamefont {Cross},\ and\ \citenamefont
  {Gambetta}}]{suchara2015}%
  \BibitemOpen
  \bibfield  {author} {\bibinfo {author} {\bibfnamefont {M.}~\bibnamefont
  {Suchara}}, \bibinfo {author} {\bibfnamefont {A.~W.}\ \bibnamefont {Cross}},\
  and\ \bibinfo {author} {\bibfnamefont {J.~M.}\ \bibnamefont {Gambetta}},\
  }\bibfield  {title} {\bibinfo {title} {{Leakage Suppression in the Toric
  Code}},\ }\href@noop {} {\bibfield  {journal} {\bibinfo  {journal} {Quantum
  Information and Computation}\ }\textbf {\bibinfo {volume} {15}},\ \bibinfo
  {pages} {997} (\bibinfo {year} {2015})},\ \bibinfo {note} {preprint version:
  2014},\ \Eprint {https://arxiv.org/abs/1410.8562} {arXiv:1410.8562
  [quant-ph]} \BibitemShut {NoStop}%
\bibitem [{\citenamefont {Camps}\ \emph {et~al.}(2024)\citenamefont {Camps},
  \citenamefont {Crawford}, \citenamefont {Geh{\'e}r}, \citenamefont
  {Gramolin}, \citenamefont {Stafford},\ and\ \citenamefont
  {Turner}}]{camps2024}%
  \BibitemOpen
  \bibfield  {author} {\bibinfo {author} {\bibfnamefont {J.}~\bibnamefont
  {Camps}}, \bibinfo {author} {\bibfnamefont {O.}~\bibnamefont {Crawford}},
  \bibinfo {author} {\bibfnamefont {G.~P.}\ \bibnamefont {Geh{\'e}r}}, \bibinfo
  {author} {\bibfnamefont {A.~V.}\ \bibnamefont {Gramolin}}, \bibinfo {author}
  {\bibfnamefont {M.~P.}\ \bibnamefont {Stafford}},\ and\ \bibinfo {author}
  {\bibfnamefont {M.}~\bibnamefont {Turner}},\ }\href@noop {} {\bibinfo {title}
  {{Leakage Mobility in Superconducting Qubits as a Leakage Reduction Unit}}}
  (\bibinfo {year} {2024}),\ \bibinfo {note} {version dated 6 June 2024},\
  \Eprint {https://arxiv.org/abs/2406.04083v1} {arXiv:2406.04083v1 [quant-ph]}
  \BibitemShut {NoStop}%
\bibitem [{\citenamefont {Manabe}\ \emph {et~al.}(2025)\citenamefont {Manabe},
  \citenamefont {Suzuki},\ and\ \citenamefont {Darmawan}}]{manabe2025}%
  \BibitemOpen
  \bibfield  {author} {\bibinfo {author} {\bibfnamefont {H.}~\bibnamefont
  {Manabe}}, \bibinfo {author} {\bibfnamefont {Y.}~\bibnamefont {Suzuki}},\
  and\ \bibinfo {author} {\bibfnamefont {A.~S.}\ \bibnamefont {Darmawan}},\
  }\href@noop {} {\bibinfo {title} {{Efficient Simulation of Leakage Errors in
  Quantum Error Correcting Codes Using Tensor Network Methods}}} (\bibinfo
  {year} {2025}),\ \bibinfo {note} {version dated 21 January 2025},\ \Eprint
  {https://arxiv.org/abs/2308.08186v2} {arXiv:2308.08186v2 [quant-ph]}
  \BibitemShut {NoStop}%
\bibitem [{\citenamefont {Marshall}\ and\ \citenamefont
  {Kafri}(2025)}]{marshall2025}%
  \BibitemOpen
  \bibfield  {author} {\bibinfo {author} {\bibfnamefont {J.}~\bibnamefont
  {Marshall}}\ and\ \bibinfo {author} {\bibfnamefont {D.}~\bibnamefont
  {Kafri}},\ }\bibfield  {title} {\bibinfo {title} {{Incoherent approximation
  of leakage in quantum error correction}},\ }\href
  {https://doi.org/10.1103/physrevapplied.23.054025} {\bibfield  {journal}
  {\bibinfo  {journal} {Physical Review Applied}\ }\textbf {\bibinfo {volume}
  {23}},\ \bibinfo {pages} {054025} (\bibinfo {year} {2025})},\ \Eprint
  {https://arxiv.org/abs/2312.10277} {arXiv:2312.10277 [quant-ph]} \BibitemShut
  {NoStop}%
\bibitem [{\citenamefont {Wu}\ \emph {et~al.}(2022)\citenamefont {Wu},
  \citenamefont {Kolkowitz}, \citenamefont {Puri},\ and\ \citenamefont
  {Thompson}}]{wu2022}%
  \BibitemOpen
  \bibfield  {author} {\bibinfo {author} {\bibfnamefont {Y.}~\bibnamefont
  {Wu}}, \bibinfo {author} {\bibfnamefont {S.}~\bibnamefont {Kolkowitz}},
  \bibinfo {author} {\bibfnamefont {S.}~\bibnamefont {Puri}},\ and\ \bibinfo
  {author} {\bibfnamefont {J.~D.}\ \bibnamefont {Thompson}},\ }\bibfield
  {title} {\bibinfo {title} {{Erasure conversion for fault-tolerant quantum
  computing in alkaline earth Rydberg atom arrays}},\ }\href
  {https://doi.org/10.1038/s41467-022-32094-6} {\bibfield  {journal} {\bibinfo
  {journal} {Nature Communications}\ }\textbf {\bibinfo {volume} {13}},\
  \bibinfo {pages} {4657} (\bibinfo {year} {2022})},\ \Eprint
  {https://arxiv.org/abs/2201.03540} {arXiv:2201.03540 [quant-ph]} \BibitemShut
  {NoStop}%
\bibitem [{\citenamefont {McEwen}\ \emph {et~al.}(2021)\citenamefont {McEwen},
  \citenamefont {Kafri}, \citenamefont {Chen}, \citenamefont {Atalaya},
  \citenamefont {Satzinger}, \citenamefont {Quintana}, \citenamefont {Klimov},
  \citenamefont {Sank}, \citenamefont {Gidney}, \citenamefont {Fowler},
  \citenamefont {Arute}, \citenamefont {Arya}, \citenamefont {Buckley},
  \citenamefont {Burkett}, \citenamefont {Bushnell}, \citenamefont {Chiaro},
  \citenamefont {Collins}, \citenamefont {Demura}, \citenamefont {Dunsworth},
  \citenamefont {Erickson}, \citenamefont {Foxen}, \citenamefont {Giustina},
  \citenamefont {Huang}, \citenamefont {Hong}, \citenamefont {Jeffrey},
  \citenamefont {Kim}, \citenamefont {Kechedzhi}, \citenamefont {Kostritsa},
  \citenamefont {Laptev}, \citenamefont {Megrant}, \citenamefont {Mi},
  \citenamefont {Mutus}, \citenamefont {Naaman}, \citenamefont {Neeley},
  \citenamefont {Neill}, \citenamefont {Niu}, \citenamefont {Paler},
  \citenamefont {Redd}, \citenamefont {Roushan}, \citenamefont {White},
  \citenamefont {Yao}, \citenamefont {Yeh}, \citenamefont {Zalcman},
  \citenamefont {Chen}, \citenamefont {Smelyanskiy}, \citenamefont {Martinis},
  \citenamefont {Neven}, \citenamefont {Kelly}, \citenamefont {Korotkov},
  \citenamefont {Petukhov},\ and\ \citenamefont {Barends}}]{mcewen2021}%
  \BibitemOpen
  \bibfield  {author} {\bibinfo {author} {\bibfnamefont {M.}~\bibnamefont
  {McEwen}}, \bibinfo {author} {\bibfnamefont {D.}~\bibnamefont {Kafri}},
  \bibinfo {author} {\bibfnamefont {Z.}~\bibnamefont {Chen}}, \bibinfo {author}
  {\bibfnamefont {J.}~\bibnamefont {Atalaya}}, \bibinfo {author} {\bibfnamefont
  {K.~J.}\ \bibnamefont {Satzinger}}, \bibinfo {author} {\bibfnamefont
  {C.}~\bibnamefont {Quintana}}, \bibinfo {author} {\bibfnamefont {P.~V.}\
  \bibnamefont {Klimov}}, \bibinfo {author} {\bibfnamefont {D.}~\bibnamefont
  {Sank}}, \bibinfo {author} {\bibfnamefont {C.}~\bibnamefont {Gidney}},
  \bibinfo {author} {\bibfnamefont {A.~G.}\ \bibnamefont {Fowler}}, \bibinfo
  {author} {\bibfnamefont {F.}~\bibnamefont {Arute}}, \bibinfo {author}
  {\bibfnamefont {K.}~\bibnamefont {Arya}}, \bibinfo {author} {\bibfnamefont
  {B.}~\bibnamefont {Buckley}}, \bibinfo {author} {\bibfnamefont
  {B.}~\bibnamefont {Burkett}}, \bibinfo {author} {\bibfnamefont
  {N.}~\bibnamefont {Bushnell}}, \bibinfo {author} {\bibfnamefont
  {B.}~\bibnamefont {Chiaro}}, \bibinfo {author} {\bibfnamefont
  {R.}~\bibnamefont {Collins}}, \bibinfo {author} {\bibfnamefont
  {S.}~\bibnamefont {Demura}}, \bibinfo {author} {\bibfnamefont
  {A.}~\bibnamefont {Dunsworth}}, \bibinfo {author} {\bibfnamefont
  {C.}~\bibnamefont {Erickson}}, \bibinfo {author} {\bibfnamefont
  {B.}~\bibnamefont {Foxen}}, \bibinfo {author} {\bibfnamefont
  {M.}~\bibnamefont {Giustina}}, \bibinfo {author} {\bibfnamefont
  {T.}~\bibnamefont {Huang}}, \bibinfo {author} {\bibfnamefont
  {S.}~\bibnamefont {Hong}}, \bibinfo {author} {\bibfnamefont {E.}~\bibnamefont
  {Jeffrey}}, \bibinfo {author} {\bibfnamefont {S.}~\bibnamefont {Kim}},
  \bibinfo {author} {\bibfnamefont {K.}~\bibnamefont {Kechedzhi}}, \bibinfo
  {author} {\bibfnamefont {F.}~\bibnamefont {Kostritsa}}, \bibinfo {author}
  {\bibfnamefont {P.}~\bibnamefont {Laptev}}, \bibinfo {author} {\bibfnamefont
  {A.}~\bibnamefont {Megrant}}, \bibinfo {author} {\bibfnamefont
  {X.}~\bibnamefont {Mi}}, \bibinfo {author} {\bibfnamefont {J.}~\bibnamefont
  {Mutus}}, \bibinfo {author} {\bibfnamefont {O.}~\bibnamefont {Naaman}},
  \bibinfo {author} {\bibfnamefont {M.}~\bibnamefont {Neeley}}, \bibinfo
  {author} {\bibfnamefont {C.}~\bibnamefont {Neill}}, \bibinfo {author}
  {\bibfnamefont {M.}~\bibnamefont {Niu}}, \bibinfo {author} {\bibfnamefont
  {A.}~\bibnamefont {Paler}}, \bibinfo {author} {\bibfnamefont
  {N.}~\bibnamefont {Redd}}, \bibinfo {author} {\bibfnamefont {P.}~\bibnamefont
  {Roushan}}, \bibinfo {author} {\bibfnamefont {T.~C.}\ \bibnamefont {White}},
  \bibinfo {author} {\bibfnamefont {J.}~\bibnamefont {Yao}}, \bibinfo {author}
  {\bibfnamefont {P.}~\bibnamefont {Yeh}}, \bibinfo {author} {\bibfnamefont
  {A.}~\bibnamefont {Zalcman}}, \bibinfo {author} {\bibfnamefont
  {Y.}~\bibnamefont {Chen}}, \bibinfo {author} {\bibfnamefont {V.~N.}\
  \bibnamefont {Smelyanskiy}}, \bibinfo {author} {\bibfnamefont {J.~M.}\
  \bibnamefont {Martinis}}, \bibinfo {author} {\bibfnamefont {H.}~\bibnamefont
  {Neven}}, \bibinfo {author} {\bibfnamefont {J.}~\bibnamefont {Kelly}},
  \bibinfo {author} {\bibfnamefont {A.~N.}\ \bibnamefont {Korotkov}}, \bibinfo
  {author} {\bibfnamefont {A.~G.}\ \bibnamefont {Petukhov}},\ and\ \bibinfo
  {author} {\bibfnamefont {R.}~\bibnamefont {Barends}},\ }\bibfield  {title}
  {\bibinfo {title} {{Removing leakage-induced correlated errors in
  superconducting quantum error correction}},\ }\href
  {https://doi.org/10.1038/s41467-021-21982-y} {\bibfield  {journal} {\bibinfo
  {journal} {Nature Communications}\ }\textbf {\bibinfo {volume} {12}},\
  \bibinfo {pages} {1761} (\bibinfo {year} {2021})}\BibitemShut {NoStop}%
\bibitem [{\citenamefont {Battistel}\ \emph {et~al.}(2021)\citenamefont
  {Battistel}, \citenamefont {Varbanov},\ and\ \citenamefont
  {Terhal}}]{battistel2021}%
  \BibitemOpen
  \bibfield  {author} {\bibinfo {author} {\bibfnamefont {F.}~\bibnamefont
  {Battistel}}, \bibinfo {author} {\bibfnamefont {B.~M.}\ \bibnamefont
  {Varbanov}},\ and\ \bibinfo {author} {\bibfnamefont {B.~M.}\ \bibnamefont
  {Terhal}},\ }\bibfield  {title} {\bibinfo {title} {{Hardware-Efficient
  Leakage-Reduction Scheme for Quantum Error Correction with Superconducting
  Transmon Qubits}},\ }\href {https://doi.org/10.1103/PRXQuantum.2.030314}
  {\bibfield  {journal} {\bibinfo  {journal} {PRX Quantum}\ }\textbf {\bibinfo
  {volume} {2}},\ \bibinfo {pages} {030314} (\bibinfo {year}
  {2021})}\BibitemShut {NoStop}%
\bibitem [{\citenamefont {van~den Berg}\ \emph {et~al.}(2023)\citenamefont
  {van~den Berg}, \citenamefont {Minev}, \citenamefont {Kandala},\ and\
  \citenamefont {Temme}}]{vandenberg2023}%
  \BibitemOpen
  \bibfield  {author} {\bibinfo {author} {\bibfnamefont {E.}~\bibnamefont
  {van~den Berg}}, \bibinfo {author} {\bibfnamefont {Z.~K.}\ \bibnamefont
  {Minev}}, \bibinfo {author} {\bibfnamefont {A.}~\bibnamefont {Kandala}},\
  and\ \bibinfo {author} {\bibfnamefont {K.}~\bibnamefont {Temme}},\ }\bibfield
   {title} {\bibinfo {title} {Probabilistic error cancellation with sparse
  {Pauli--Lindblad} models on noisy quantum processors},\ }\href
  {https://doi.org/10.1038/s41567-023-02042-2} {\bibfield  {journal} {\bibinfo
  {journal} {Nature Physics}\ }\textbf {\bibinfo {volume} {19}},\ \bibinfo
  {pages} {1116} (\bibinfo {year} {2023})},\ \Eprint
  {https://arxiv.org/abs/2201.09866} {arXiv:2201.09866 [quant-ph]} \BibitemShut
  {NoStop}%
\bibitem [{\citenamefont {Hines}\ \emph {et~al.}(2026)\citenamefont {Hines},
  \citenamefont {Ostrove}, \citenamefont {Rudinger}, \citenamefont {Seritan},
  \citenamefont {Young}, \citenamefont {Blume-Kohout},\ and\ \citenamefont
  {Proctor}}]{hines2026}%
  \BibitemOpen
  \bibfield  {author} {\bibinfo {author} {\bibfnamefont {J.}~\bibnamefont
  {Hines}}, \bibinfo {author} {\bibfnamefont {C.}~\bibnamefont {Ostrove}},
  \bibinfo {author} {\bibfnamefont {K.}~\bibnamefont {Rudinger}}, \bibinfo
  {author} {\bibfnamefont {S.}~\bibnamefont {Seritan}}, \bibinfo {author}
  {\bibfnamefont {K.}~\bibnamefont {Young}}, \bibinfo {author} {\bibfnamefont
  {R.}~\bibnamefont {Blume-Kohout}},\ and\ \bibinfo {author} {\bibfnamefont
  {T.}~\bibnamefont {Proctor}},\ }\href@noop {} {\bibinfo {title} {{Simulating
  Quantum Error Correction beyond Pauli Stochastic Errors}}} (\bibinfo {year}
  {2026}),\ \bibinfo {note} {preprint},\ \Eprint
  {https://arxiv.org/abs/2603.18457v1} {arXiv:2603.18457v1 [quant-ph]}
  \BibitemShut {NoStop}%
\bibitem [{\citenamefont {Choi}(1975)}]{choi1975}%
  \BibitemOpen
  \bibfield  {author} {\bibinfo {author} {\bibfnamefont {M.-D.}\ \bibnamefont
  {Choi}},\ }\bibfield  {title} {\bibinfo {title} {{Completely positive linear
  maps on complex matrices}},\ }\href
  {https://doi.org/10.1016/0024-3795(75)90075-0} {\bibfield  {journal}
  {\bibinfo  {journal} {Linear Algebra and its Applications}\ }\textbf
  {\bibinfo {volume} {10}},\ \bibinfo {pages} {285} (\bibinfo {year}
  {1975})}\BibitemShut {NoStop}%
\bibitem [{\citenamefont {Higham}(2008)}]{higham2008}%
  \BibitemOpen
  \bibfield  {author} {\bibinfo {author} {\bibfnamefont {N.~J.}\ \bibnamefont
  {Higham}},\ }\href {https://doi.org/10.1137/1.9780898717778} {\emph {\bibinfo
  {title} {Functions of Matrices: Theory and Computation}}}\ (\bibinfo
  {publisher} {Society for Industrial and Applied Mathematics},\ \bibinfo
  {year} {2008})\BibitemShut {NoStop}%
\bibitem [{\citenamefont {Schmeding}(2024)}]{schmeding2024lie}%
  \BibitemOpen
  \bibfield  {author} {\bibinfo {author} {\bibfnamefont {A.}~\bibnamefont
  {Schmeding}},\ }\href
  {https://wiki.math.ntnu.no/_media/ma3407/2024h/introduction_to_lie_theory.pdf?rev=1731658966}
  {\bibinfo {title} {Notes {MA3407}: Introduction to {Lie} theory}} (\bibinfo
  {year} {2024}),\ \bibinfo {note} {nTNU lecture notes, draft version
  1.6}\BibitemShut {NoStop}%
\bibitem [{\citenamefont {Wolf}\ and\ \citenamefont
  {Cirac}(2008)}]{wolfcirac2008}%
  \BibitemOpen
  \bibfield  {author} {\bibinfo {author} {\bibfnamefont {M.~M.}\ \bibnamefont
  {Wolf}}\ and\ \bibinfo {author} {\bibfnamefont {J.~I.}\ \bibnamefont
  {Cirac}},\ }\bibfield  {title} {\bibinfo {title} {Dividing quantum
  channels},\ }\href {https://doi.org/10.1007/s00220-008-0411-y} {\bibfield
  {journal} {\bibinfo  {journal} {Communications in Mathematical Physics}\
  }\textbf {\bibinfo {volume} {279}},\ \bibinfo {pages} {147} (\bibinfo {year}
  {2008})},\ \Eprint {https://arxiv.org/abs/math-ph/0611057}
  {arXiv:math-ph/0611057} \BibitemShut {NoStop}%
\bibitem [{\citenamefont {Bacon}\ \emph {et~al.}(2001)\citenamefont {Bacon},
  \citenamefont {Childs}, \citenamefont {Chuang}, \citenamefont {Kempe},
  \citenamefont {Leung},\ and\ \citenamefont {Zhou}}]{bacon2001}%
  \BibitemOpen
  \bibfield  {author} {\bibinfo {author} {\bibfnamefont {D.}~\bibnamefont
  {Bacon}}, \bibinfo {author} {\bibfnamefont {A.~M.}\ \bibnamefont {Childs}},
  \bibinfo {author} {\bibfnamefont {I.~L.}\ \bibnamefont {Chuang}}, \bibinfo
  {author} {\bibfnamefont {J.}~\bibnamefont {Kempe}}, \bibinfo {author}
  {\bibfnamefont {D.~W.}\ \bibnamefont {Leung}},\ and\ \bibinfo {author}
  {\bibfnamefont {X.}~\bibnamefont {Zhou}},\ }\bibfield  {title} {\bibinfo
  {title} {Universal simulation of {Markovian} quantum dynamics},\ }\href
  {https://doi.org/10.1103/PhysRevA.64.062302} {\bibfield  {journal} {\bibinfo
  {journal} {Physical Review A}\ }\textbf {\bibinfo {volume} {64}},\ \bibinfo
  {pages} {062302} (\bibinfo {year} {2001})},\ \Eprint
  {https://arxiv.org/abs/quant-ph/0008070} {arXiv:quant-ph/0008070 [quant-ph]}
  \BibitemShut {NoStop}%
\bibitem [{\citenamefont {Sweke}\ \emph {et~al.}(2015)\citenamefont {Sweke},
  \citenamefont {Sinayskiy}, \citenamefont {Bernard},\ and\ \citenamefont
  {Petruccione}}]{sweke2015}%
  \BibitemOpen
  \bibfield  {author} {\bibinfo {author} {\bibfnamefont {R.}~\bibnamefont
  {Sweke}}, \bibinfo {author} {\bibfnamefont {I.}~\bibnamefont {Sinayskiy}},
  \bibinfo {author} {\bibfnamefont {D.}~\bibnamefont {Bernard}},\ and\ \bibinfo
  {author} {\bibfnamefont {F.}~\bibnamefont {Petruccione}},\ }\bibfield
  {title} {\bibinfo {title} {Universal simulation of {Markovian} open quantum
  systems},\ }\href {https://doi.org/10.1103/PhysRevA.91.062308} {\bibfield
  {journal} {\bibinfo  {journal} {Physical Review A}\ }\textbf {\bibinfo
  {volume} {91}},\ \bibinfo {pages} {062308} (\bibinfo {year} {2015})},\
  \Eprint {https://arxiv.org/abs/1503.05028} {arXiv:1503.05028 [quant-ph]}
  \BibitemShut {NoStop}%
\end{thebibliography}
\end{document}